# Why Transaction Cost Economics Failed and How to Fix It

Mingqian Li

（School of Economics， Southwestern University of Finance and Economics， Chengdu）

**Abstract**

The connotation of transaction costs has never been definitively determined, and the independence of the concept has never been rigorously demonstrated. This paper delves into the thought systems of several prominent economists in the development of transaction cost economics, starting from first-hand materials. By combining multiple works of the authors, it reconstructs the true meanings and identifies endogeneity issues and logical inconsistencies. The conclusion of this paper is bold. Previous research has been largely filled with misinterpretations and misunderstandings, as people have focused solely on the wording of transaction cost definitions, neglecting the nature of transaction costs. The intention of transaction cost theory has been unwittingly assimilated into the objects it intends to criticize. After delineating the framework of "transaction costs-property rights-competition", this paper reconstructs the concept of transaction costs and the history of transaction cost concepts, providing a direct response to this theoretical puzzle that has plagued the academic community for nearly a century.

## 1 Introduction

In his seminal work "The Nature of the Firm", Coase began with the following critique:

> "Economic theory has suffered in the past from a failure to state clearly its assumptions. Economists in building up a theory have often omitted to examine the foundations on which it was erected. This examination is, however, essential not only to prevent the misunderstanding and needless controversy which arise from a lack of knowledge of the assumptions on which a theory is based, but also because of the extreme importance for economics of good judgment in choosing between rival sets of assumptions." (Coase,1990, p.33)

However, the paper arrow shot by Coase ultimately landed on his own nose— as the founder of new institutional economics, Coase did not witness a convincing explanation of the transaction costs he introduced. Instead, they remained acknowledged but undiscussed.

The ambiguity of fundamental concepts did not impede the progress of New Institutional Economists. Through the efforts of several generations of scholars, economists refined the theory of the firm, and developed contract theory, property rights theory, and institutional economics based on the cornerstone of transaction costs. However, the "transaction cost revolution" was not smooth sailing. Due to a lack of necessary discussion on basic concepts, the credibility of the theory was significantly undermined. Calls to discard transaction cost theory were persistent, and there was a pervasive distrust of the transaction cost paradigm even among institutional economists. Fischer (1977, p.322) bluntly commented that transaction cost economics has garnered "a well-deserved bad name". Dietrich (2008, p.14) stated: "Fischer's criticism is correct".

The core debates surrounding the concept of transaction costs mainly include two points: (1) transaction costs lack a clear and universally recognized definition; (2) transaction costs are difficult to distinguish from production costs. The purpose of this paper is to revisit the theoretical achievements of predecessors, elucidate the nature of transaction costs based on the frameworks established by the scholars, and attempt to directly address this theoretical challenge that has been shelved for decades.

It should be noted that the reason why I use many direct citations and avoid citing review literature as much as possible is precisely because the academic community has long focused only on second-hand materials, without anyone returning to the original text and improving the transaction cost theory from the author's original intention.[1]

---

[1] This paragraph wasn't originally included. After having my work rejected by multiple journals and even being refused for review, I added a number of sharp statements. If you open books on the history of philosophy or political philosophy, you'll find that no one avoid using firsthand materials. However, for researchers of economic

## 2 History of Transaction Cost Concept

After decades of research,[2] it seems that there is now enough confidence to provide clear answers to some important questions. Why do firms exist?[3] Why do legal systems emerge? Why are markets efficient? Because transaction costs are positive.[4] However, what are transaction costs? Are they simply the costs of exchange? If transaction costs are only related to exchange activities, why are they used to explain the emergence of institutions? Can all issues be attributed to transaction costs? How do we determine the boundaries of transaction costs? How do we distinguish between transaction costs and production costs? Although the concept of transaction costs has been applied extensively in various aspects of economic theory, economists still have not settled on a definitive understanding of what transaction costs entail, which is an extremely rare phenomenon in any discipline and should not occur.

Indeed, each scholar who defined transaction costs focused on different aspects and aimed to solve different problems, making it reasonable for their views to vary in content and expression. However, if

---

thoughts, this is unimaginable. Some have deemed my article to be of insufficient quality simply because I didn't cite a well-known foreign scholar's review paper. But the problem is, what I am criticizing is exactly the firsthand materials these secondary studies rely on. It's like criticizing a chef for adding Pepsi instead of water when making soup, yet he claims to be a Michelin three-star chef and that others love his cooking. Ask yourself: if the firsthand materials are flawed, what is the use of secondary research?

[2] These paragraphs were not originally included. I added them to address the criticism from some scholars who questioned the significance of my research on the history of transaction cost concepts. Interestingly, the reason why I communicated with some of these scholars is precisely because I have read their previous papers discussing the definition of transaction costs.

[3] This paper aims to fully reflect my thought process through rhetorical and interrogative sentences, helping readers form and understand the research approach in the history of economic concepts. The basic framework for the study of the history of economic concepts includes three points: first, what problem are we trying to solve through this concept? Second, what is "the way we think" behind this concept? Third, does this concept possess independence?

[4] This is the most absurd statement I've ever read—why is the market efficient? Because market transactions incur costs. It's like asking someone, "Why do you walk to work?" and they reply, "Because walking consumes energy." This is a typical example of economists completely have no idea what they're talking about. Please don't forget, Coase's seminal work "The Nature of the Firm" begin precisely by questioning the effectiveness of an entirely free market!

a thorough reading and independent analysis of each scholar's theoretical framework reveals inconsistencies in logic, or serious contradictions between their conceptual discussions and practical applications, it becomes necessary to revisit existing definitions and seek new solutions. Transaction costs are a cornerstone of modern Western economic theory, and in Marxist political economic theory, "circulation costs" are used in place of "transaction costs". If this concept loses its value and fails to fulfill its purpose, the validity of the entire theoretical structure should be questioned. Therefore, this paper does not adopt a method of horizontal comparison among different views but delves into the individual thought systems of each author to identify inherent problems and logical inconsistencies.[5]

Unfortunately, after revisiting the history of the concept of transaction costs, we arrive at a disappointing conclusion: Although there was a hope to use the concept of transaction costs to innovate and break away from traditional methods such as the Walrasian system and "blackboard economics", the independence of the transaction cost concept was achieved by narrowly defining the concept of production costs. This has led to the assimilation of transaction cost theory into traditional methods. Consequently, the various theories built on the traditional concept of transaction costs have inevitably become illusory, like the moon reflected in water or flowers in a mirror. The cause of this predicament is the accumulation of numerous misunderstandings and misinterpretations in past research.

The logic of reality does not necessarily adhere to the historical sequence or judgments of merit and demerit. [6]Therefore, the analysis of the views of various scholars in this paper is not conducted strictly in the order in which the ideas were conceived, but rather unfolds according to the mainstream academic community's gradual understanding of transaction costs.

---

[5] This passage is a response to scholars who, without reading my article, have already labeled it as a hodgepodge of traditional theories. However, in the very early stages of this article's conception (the outline of "The Market Nature of Transaction Costs"), I did attempt to blend together a garbage definition.

[6] The study of the history of thoughts is not about awarding credit or recognition, and the work of different scholars does not necessarily contribute to the same theoretical framework. There is no need to forcefully unite everyone under one banner. History of thoughts becomes meaningful only when it restores the development of scholarly understanding of a particular idea, uncovering the emergence, decline, and elevation of the ways people thought at that time.

### 2.1 Two Discoveries of Transaction Costs

My argument is quite simple. First, I will prove that Arrow served transaction cost as market failure and externality. Second, I will prove that Coase's transaction cost is neither market failure nor externality. Thus, Arrow's view on transaction costs does not have reference value.

#### 2.1.1 Arrow's transaction cost: The biggest misunderstanding in the history of economic concepts

A debate sparked by Arrow illustrated the gradual expansion of transaction cost thinking into the mainstream perspective (Klaes,2000, p.572). Common understanding holds that this debate was a clash between the transaction cost perspective and general equilibrium theory, with transaction cost theory emerging triumphant. The definition of transaction costs by Arrow has been widely referenced and adopted, becoming a pivotal pillar in the current construction of the transaction cost concept. Contrary to popular belief, however, the outcome of the debate was actually a tragedy for transaction cost theory—it ended up being misunderstood as the very issues it intended to critique, namely "externalities" and "market failures" —yet this misinterpretation remained unnoticed and misled countless scholars.[7] What adds a tragic dimension is the widespread dissemination of Arrow's ideas, which proves that most people are unaware of the opposition between transaction costs and concepts like externalities and market failure, failing to recognize this as a tragedy. In what follows, we will demystify this historic debate to understand how transaction costs re-entered the economist's discourse and how they have been misunderstood.

In his study of the absence of medical insurance markets, Arrow (1963, p.945) suggested that to achieve optimum results, the government should employ compulsive measures such as subsidies and taxes to increase the supply of insurance. On the other side of the debate, Lees and Rice (1965, p.140) disagreed with Arrow's call for government intervention. Up to this point, this story is merely a microcosm of the broader mid-20th century debate on the relationship between government and the market. However, what distinguishes their discussion is their utilization of the concept of transaction costs.

[7] I naively believed that revealing Arrow's transaction cost was referring to externalities would be enough to make others take this paper seriously. However, in discussions with other people, I was surprised to find that none of them knew that transaction costs and externalities are not the same thing. What I thought was common knowledge now requires me to spend a considerable amount of effort explaining the differences between these two ways of thinking.

Lees and Rice did not explore the deeper implications of transaction costs; they seemingly equated transaction costs with the literal meaning of transaction costs. Specifically, in the context of the medical insurance market, they defined these as "the transaction costs to the individual of completing and filing applications and claims forms, paying premiums, keeping records, etc., as well as possible costs of obtaining information" (Lees & Rice, 1965, p.143). They argued that Arrow had failed to consider these costs, which "may be of sufficient magnitude to make insurance policies against certain losses not worthwhile" (Lees & Rice, 1965, p.143), thus challenging Arrow's conclusions.

Let's now examine how Arrow responded to the criticisms of the two scholars. It is important to note that the theme of this response remains the failure of government and markets. Arrow (1965, pp.155-156) acknowledged the existence of transaction costs but disagreed with the conclusions of the two scholars, arguing that there was no convincing argument to prove that markets save more on transaction costs than governments.

It is worth mentioning that in this response, Arrow also offered his preliminary understanding of transaction costs:

> "The transaction cost may represent a real cost independent of the method of market organization, for example, transportation cost…...Alternatively, transaction costs may inhere in a particular mode of economic organization, and may be avoided by switching to a different system." (Arrow,1965, p.155)

Arrow simply categorized transaction costs—initially suggesting that they "may" be independent of organizational structure, and later noting that they "may" be closely associated with it. Apparently, this assertion offers almost no value in understanding the nature of transaction costs—first, because Arrow's expression is quite vague; second, it remains unclear why costs unrelated to organization can be considered transaction costs, and why those closely related to organizational structure can also be classified as such. We must inquire further: what exactly unifies these seemingly unrelated costs, and what are the qualitative factors behind each of them? However, Arrow (1965, pp.155-158) did not explore this further, and in the subsequent sections of his paper, he focused only on discussing and applying transaction costs related to organizational systems.

If the use of terms such as "may" and "alternatively" indicated Arrow's initial hesitation, by the time he wrote his seminal 1969 article, it is clear that he had undertaken systematic reflection. Arrow claimed:

> "Current writing has helped bring out the point that market failure is not absolute; it is better to consider a broader category that of transaction costs, which in general impede and in particular cases completely block the formation of markets. It is usually though not always emphasized that transaction costs are costs of running the economic system." (Arrow,1969, p.48)

This statement has been widely referenced and cited, becoming an essential definition that any scholar attempting to discuss the definition of transaction costs cannot ignore.[8] From the semantics of this passage, it seems that Arrow sought to reformulate the traditional view of market failure through the lens of transaction costs. Next, we will unveil the true connotation behind Arrow's statement.

In the section on "Transaction Costs", Arrow identified three sources of transaction costs:

> "The discussions in the preceding sections suggest two sources of transaction costs. (1) exclusion costs; (2) costs of communication and information, including both the supplying and the learning

[8] If I were to define transaction costs, I would say "the cost of organizing resources (elements)". From the literal meaning, there is not much difference between it and Arrow's definition. But from the perspective of "the way we think", the two are completely opposite. This is exactly the core point of my paper—do not focus on form, but on substance; do not focus on the wording, but the way of viewing the world.

> of the terms on which transactions can be carried out. An additional source is (3) the costs of disequilibrium; in any complex system, the market or authoritative allocation, even under perfect information, it takes time to compute the optimal allocation, and either transactions take place which are inconsistent with the final equilibrium or they are delayed until the computations are completed." (Arrow,1969, p.60)

Similarly, as we further investigate "the discussions in the preceding sections",[9] we find that Arrow also identified three causes for market failure:

> "Previous discussion has suggested two possible causes for market failure (1) inability to exclude;(2) lack of necessary information to permit mark transactions to be concluded. The failure of futures markets cannot be directly explained in the terms……The absence of futures markets may be ascribed to a third possibility (3) supply and demand are equated at zero; the highest price at which anyone would buy is below the lowest price at which anyone would sell. This third case of market failure, unlike the first two, is by itself in no way presumptive of inefficiency. However, it may usually be assumed that its occurrence is the result of failures of the first two types on complementary markets. Specifically, the demand for future steel may be low because of uncertainties of all types; sales and technological uncertainty for the buyer's firm, prices and existence of competing goods, and the quality specification of the steel. If, however, adequate markets for risk-bearing exist, the uncertainties could be removed, and the demand for future steel would rise." (Arrow,1969, p.59)

---

[9] Although I criticized scholars who enjoy skipping, I also used skipping when studying Arrow's thought. There are two reasons that ultimately set me apart from these individuals. Firstly, the analysis of the history of economy concept in this paper was added later. Before these analyses, I had already reached the "New Trinity". After realizing that I had made innovations that could overturn Nobel laureates' theory, I naturally had the patience to force myself to make the paper solid. Secondly, I did not initially understand Arrow's opinion about the third source of transaction costs, and I believe most people don't understand either. However, I had learned in primary school that when you don't understand a paragraph, you should look over the surrounding context. The paragraph order of each section in this article roughly reflects my reading sequence, reflecting my thoughts and emotional journey. That's why readers may have a sense of turning the tide and uncovering a big lie when reading this paper. At first, I read the paper out of utilitarian motives, but when I realized that Arrow had made a grave mistake and no one had noticed, my attitude shifted from pride to anger.

Despite Arrow's proclivity for verbose language, we can succinctly summarize his concepts of "the costs of disequilibrium" and "supply and demand are equated at zero" as follows: market prices deviate from equilibrium prices due to various reasons ("equilibration being a process" or "uncertainty"), resulting in welfare losses.

After comparing the three sources of transaction costs with the three causes of market failure, we arrive at a startling discovery: under Arrow's analysis, the three sources of transaction costs correspond directly to the three causes for market failure. Apparently, Arrow equated "transaction costs" with "market failure". Additionally, Arrow employed the analytical framework of externalities when discussing the impact of transaction costs: "In a price system, transaction costs drive a wedge between buyer's and seller's prices and thereby give rise to welfare losses as in the usual analysis." (Arrow,1969, p.60) It is evident that Arrow equated "transaction costs" with "externalities". Arrow also noted: "The problem of externalities is thus a special case of a more general phenomenon, the failure of markets to exist." (Arrow,1969, p.59) Regarding the concept of the "economic system" in the definition, Wang claimed that since Arrow did not clarify the specific meaning of this term, Arrow's definition of transaction costs lacks operability (Wang, 1996, p.42). In fact, "economic system" refers to the forms of resource allocation such as markets and government (Arrow, 1969, p.60). We should not conclude that Arrow failed to explain its implications clearly simply because he did not provide a precise definition; a glance at the title of Arrow's article is sufficient to grasp his true intent.[10]

Synthesizing all the content above, we can draw the following conclusion: Arrow's transaction costs— "costs of running the economic system" — for market-based organizational structures, result in welfare losses due to externalities and market failures. Under certain conditions, these costs can be mitigated by the visible hand of the government (at the expense of government transaction costs, or compulsion), which are quantitatively lower than the welfare loss, thus enabling savings in market transaction costs.

---

[10] Although I criticized Wang Dingding here, anyone who has read his paper will recognize his remarkable genius—during an era of severely underdeveloped information, he was just a step away from uncovering "the cost of maximizing oneself".

This assertion seems unproblematic, but when we remove the mask of "transaction costs" and revert it to the discourse of "externalities—market failure", we effortlessly recognize that Arrow is conveying a very familiar conclusion, one that swept through the economics community in the last century and remains vigorous today—government intervention can resolve externalities and market failures.

Furthermore, in terms of core issue, Arrow's three articles consistently focused on externalities and market failures. In terms of content, while Arrow appears to analyze transaction costs, in essence, the discussion revolves around externalities and market failures. Methodologically, Arrow acknowledged the use of welfare economics approach in these three articles,[11] a method that stands as a critique target of transaction cost economics. In conclusion, whether subjectively or objectively considered, this debate does not revolve around the clash between the perspective of transaction costs and general equilibrium theory. Arrow simply dresses up externalities and market failures under the guise of "transaction costs"—akin to the tale of "The Emperor's New Clothes".[12]

### 2.1.2 Coase's Transaction Costs

In stark contrast to Arrow's equation of "transaction costs" with "externalities" and "market failures", Coase himself viewed the widespread dissemination of the concept of externalities in economics as an "unfortunate" occurrence (Coase,1990, p.26). He endeavors to avoid associating his own work with "externalities." However, the concept of externalities persists like a ghost, much to his regret:

> "To prevent it being thought that I shared the common view, I never used the word 'externality' in 'The Problem of Social Cost', but spoke of "harmful effects" without specifying whether decision-makers took them into account or not……However, I was clearly unsuccessful in cutting my argument loose from the dominant approach, since 'The Problem of Social Cost' is often described, even by those sympathetic to my point of view, as a study of the problem of

---

[11] In the first two articles, Arrow explicitly stated his adoption of the welfare economics approach (see Arrow, 1963, p.941; 1965, p.156). However, in the third article, Arrow's primary aim was to clarify the meaning of the term "externalities" (see Arrow, 1969, p.47).

[12] Arrow's mistake is understandable, but the scholars who followed him are truly culpable—in order to wear the emperor's new clothes, some people even took off their own trousers!

‘externality’.” (Coase,1990, p.27)

Coase’s seminal work directly relevant to the core issue of this debate is the renowned “The Problem of Social Cost,” which is an extension of his earlier work, “The Federal Communications Commission” (Coase, 1990, p.10). Before determining the nature of transaction costs, we must address fundamental questions: what is the role and purpose of the concept of transaction costs? What problems do we aim to solve using transaction costs? Therefore, to comprehend the essence of transaction costs, we must trace back to how Coase originally discovered them.

#### 2.1.2.1 The transaction costs in “The Nature of the Firm”

At the age of 27, Coase (1990, pp.36-37) posed a crucial question in “The Nature of the Firm” that traditional theory struggled to answer: If the price mechanism is indeed all-powerful, why is it replaced by firms?[13] It was from this question that Coase recognized the existence of a certain cost, later termed “transaction costs.” Coase introduced transaction costs with the primary aim of explaining the emergence of firms. However, he neither provided a definition of transaction costs nor a definition of firms from the perspective of transaction costs in the text. Hence, to grasp Coase’s true implications of transaction costs, it is essential to examine the image of firms as portrayed by Coase.

[13] Here is the origin of my paper. At the time, I was a freshman, eagerly hoping for the opportunity to secure an academic recommendation for graduate school. I’m not the type of person who are good at mathematic, so I decided to read theoretical, text-based papers to estimate the writing difficulty, hoping for write one myself. I read many journal papers in political economics, dismissing them as something I could easily achieve. In fact, I could write one, but at that time, I didn’t know the rules of the academic game—given my background, even if I wrote something, I still wouldn’t be able to get it published. In our microeconomics class, Professor Li Yi recommended “The Nature of the Firm” to us, saying there were no mathematical formulas and no difficulties in understanding. Meanwhile, I had failed a higher algebra exam after spending nearly half the time in the restroom due to diarrhea (spicy food). By the time I came out, my legs were weak. After failing the exam, the only path to academic recommendation was through publishing papers, so I reluctantly began reading these papers. The first time I read them, I found them unremarkable and nearly fell asleep. But after a while, I began to think about how Coase discovered transaction costs, because only by doing so could I write a similar paper to secure my graduate school recommendation. Looking back now, this is the origin of my current theory—don’t focus on what is visible, but rather discover what’s right in front of us, yet intangible and invisible; don’t think about phenomena—that’s what second-rate economists do—think about how to think!

Coase argued that the commonly accepted notion in economics of “resources allocating themselves according to prices” does not apply within firms. The price mechanism operates independently without supervision, while the operation of firms requires administrative control, whereas the “invisible hand” is absent (Coase, 1990, pp.35-37). Coase stated:

> “It is true that contracts are not eliminated when there is a firm, but they are greatly reduced......For this series of contracts is substituted one……The contract is one whereby the factor, for a certain remuneration (which may be fixed or fluctuating) agrees to obey the directions of an entrepreneur within certain limits. The essence of the contract is that it should only state the limits to the powers of the entrepreneur. Within these limits, he can therefore direct the other factors of production.” (Coase,1990, p.39)

This assertion involves two key propositions: firstly, the essence of a firm lies in factors markets replacing product markets through contracts, wherein factors of production partially relinquish rights and accept the entrepreneur’s directives to gain profit; [14]secondly, the close connection between firms and contracts,[15] both serving as constraints on individuals in a completely free market, with constraints existing both on entrepreneurs and on factors of production and their owners within firms.

In conclusion, firms are organizations that correspond to markets, where if the characteristic of markets is “freedom,” then the characteristic of firms is “constraint.” These constraints are established to economize on the costs present in the market when certain constraints are absent. In formal terms, Coase (1990, p.38; p.39; p.45) utilized the expression of costs related to the costs of price mechanisms, contracting costs, market operation costs, and organizational costs.

#### 2.1.2.2 The Transaction Costs in “The Problem of Social Cost”

Coase’s discussion on transaction costs did not cease; rather, it was continued in “The Problem of

---

[14] The statement is based on Cheung’s interpretation of “The Nature of the Firm” (see Cheung,2005, pp.204-214).

[15] Cheung has revised Coase’s viewpoint, suggesting that firms are a form of contractual arrangement (see Cheung,2005, p.204), and the process of firms substituting for markets essentially involves one contract replacing another (see Cheung,2005, p.214).

Social Cost.” Coase later concluded:

> “I showed in ‘The Nature of the Firm’ that, in the absence of transaction costs, there is no economic basis for the existence of the firm. What I showed in ‘The Problem of Social Cost’ was that, in the absence of transaction costs, it does not matter what the law is, since people can always negotiate without cost to acquire, subdivide. In such a world the institutions which make up the economic system have neither substance nor purpose.” (Coase,1990, p.14)

Coase’s formulation has raised new questions: With the expanded scope of transaction cost interpretation, has there been a change in the scope and meaning of transaction costs? Coase provides several examples related to transaction costs:

> “In order to carry out a market transaction, it is necessary to discover who it is that one wishes to deal with, to inform people that one wishes to deal and on what terms.to conduct negotiations leading up to a bargain, to draw up the contract, to undertake the inspection needed to make sure that the terms of the contract are being observed, and so on. These operations are often extremely costly, sufficiently costly at any rate to prevent many transactions that would be carried out in a world in which the pricing system worked without cost.” (Coase,1990, p.114)

It is evident that this cost is closely associated with contracts and is a factor beyond traditional analysis. Based on all the above, we can conclude that Coase’s transaction costs in practical application are indeed its literal meaning—the costs of exchange. This assertion finds direct confirmation in another article by Coase (Coase,1998, p.73).

Coase did not provide a formal definition of transaction costs, but he endowed it with new significance, leading to what is now famously known as the Coase Theorem.[16] To comprehend the Coase Theorem

---

[16] I firmly believe that if you cannot express your argument clearly in one sentence, it means you haven’t thought it through. Now, I will demonstrate the power of explaining things in one sentence: People commonly believe that the Coase Theorem means the market automatically achieves Pareto efficiency when transaction costs are zero; however, the true meaning of the Coase Theorem is that the market (present situation) is always Pareto efficient, the reason you think it’s not is that there are some costs you haven’t considered, and these costs are transaction

accurately, one must consider the viewpoints of its "chief expositor" Steven N. S. Cheung.[17] According to Cheung, the Coase Theorem has three versions. Coase provided the first version in "The Federal Communications Commission," stating that "the delineation of rights is an essential prelude to market transactions" (Cheung,2005, p.250). In "The Problem of Social Cost," Coase presented the remaining two versions. One is "if property rights are clearly delineated and if all costs of transactions are zero, then resource use will be the same regardless of who owns the property rights" (Cheung, 2005, p.250). However, Coase later overturned this assertion personally, under Cheung's correction (Coase, 1990, p.15). Nonetheless, this conclusion has become a fundamental principle in economics, impervious even to Coase's own refutation. The other version states "if rights are clearly delineated and transaction costs zero, then the Pareto condition (or economic efficiency) will be satisfied" (Cheung, 2005, p.251). Cheung went further, suggesting that the so-called inefficiency is due to insufficiently comprehensive considerations (Cheung, 2019a, p.58).

As we can see, all three versions of the Coase Theorem implicitly or explicitly link "transaction costs" with "property rights". In the first formulation, Coase clearly indicated the decisive role of well-defined property rights in reducing transaction costs, and he (1990, p.13) later emphasized that "transaction costs therefore play a crucial role in determining how rights will be used." Thus, in Coase's theoretical framework, transaction costs and property rights are mutually determining, essentially two sides of the same coin.

And what role does transaction costs play in the functioning of the economic system? Coase (1990, p.14) indicated his agreement with portraying transaction costs as "friction," by quoting Stigler. It is widely understood that transaction costs act as "frictional forces" in the economic world, hindering the smooth operation of markets. Let's delve a bit deeper.

---

costs.

[17] I really can't understand that given the fact that Coase once entrusted Cheung to explain the "Coase Theorem" to others, why do some people still do not accept Cheung's interpretation?

Coase claimed that in a zero-transaction cost world, “the institutions which make up the economic system have neither substance nor purpose.” In other words, disregarding transaction costs means disregarding the practical foundation of economic institutions. Coase went on to write:

> “It would not seem worthwhile to spend much time investigating the properties of such a world. What my argument does suggest is the need to introduce positive transaction costs explicitly into economic analysis so that we can study the world that exists……However, it is my opinion that we will not be able to do this unless we first discard the approach at present used by most economists.” (Coase,1990, pp.15-16)

Undoubtedly, “such a world” refers to a zero-transaction cost world, and “the approach at present used by most economists” refer to general equilibrium methods. In general equilibrium methods, only supply and demand are core factors, while in real life, there are countless factors influencing human behavior. In other words, Coase believed that describing people’s choices and behaviors with just a few curves or equations is an oversimplification, much like envisioning a completely smooth plane and disregarding the objective existence of friction in physical analysis. The parts that should not be omitted are transaction costs, which are the “substance” and “purpose” mentioned earlier. Since they are “substance” and “purpose” of economic system, why should we call them frictional “forces” with obstructive effects? In fact, the notion of obstruction only exists in the analytical methods criticized by Coase.

Coase appropriated Stigler’s discourse on “friction”, directing criticism towards the implicit and oversimplified assumption of “zero transaction cost” in the conventional economic theory. He merely emphasized the excessive simplification of traditional methods regarding transaction costs as a foundation of real world, there is no evidence to suggest that Coase’s transaction costs hinder the operation of society. Regrettably, Coase did not specify what can be simplified and what cannot. In conclusion, interpreting transaction costs as obstacles to market operation based solely on Coase’s analogy is undoubtedly taking his words out of context and misinterpreting his intention.

### 2.1.3 Summary

We now know that transaction costs are neither externalities nor market failures. Ironically, despite Coase's opposition to interpreting his ideas through the perspective of externalities, it was the externalities that has brought transaction costs back into the economists' spotlight after decades of neglect.

Coase's discourse involves two perspectives.[18] The first emphasizes that in a completely free market, the cost of discovering prices is prohibitively high, thus negating the existence of a completely free market. The second perspective underscores the oversimplification resulting from neglecting real costs. Coase assigned two tasks to the concept of transaction costs: the first is to explain why constraints on markets are necessary, and the second is to observe various costs in the real world, with "exchange costs" being just the beginning (Coase, 1990, p.73). In conclusion, an appropriate definition of transaction costs must encompass both perspectives and fulfill both tasks; merely stating that transactions involve costs is evidently insufficient.

Now, we can clearly discern the divergence between Arrow and Coase's perspectives. Arrow interpreted transaction costs from the standpoint of externalities and market failure, contending that transaction costs cause externalities and market failures, thus obstructing market operation. However, Coase aimed to demonstrate that the existence of transaction costs is neutral. It is not transaction costs that cause externalities, but rather, due to the presence of transaction costs, the rational choice of individuals is not to address externalities. Coase stated:

> "As already indicated, the only reason individuals and private organizations do not eliminate them is that the gain from doing so would be offset by what would be lost (including the costs of making the arrangements necessary to bring about this result)." (Coase,1990, p.27)

---

[18] Originally, the structure of this article did not have a clear through-line. The initial design was to list the ideas of scholars that had to be mentioned, followed by my own ideas, allowing readers to compare them. My first reader was my good friend Xu Hanliang, and the draft of the article was still incomplete at the time. Xu Hanliang suggested that I should consider both horizontal comparisons and vertical continuity. It was based on his suggestion that I began thinking about the issue of structure and ultimately uncovered these two perspectives.

Therefore, the logic of "externalities" focuses solely on the effects without seeking underlying causes, whereas transaction costs represent the reasons for these effects.

Is it feasible to forcefully link externalities and transaction costs together? Logically, it might seem plausible, but practically, it is not. [19] The reason lies in the practical implications: externalities entail one party compensating another, whereas Coase's conception of transaction costs suggests a preliminary examination of the issue before discussing compensation—determining whether the impact truly exists or is merely subjectively perceived by the analyst. In other words, merely pointing out the existence of externalities—that is, the impact of agent A on non-decision-making agent B—is far from sufficient. Without considering the reasons, causes, and methods of addressing externalities, the concept of externalities remains an "empty shell." However, once these factors (transaction costs) are considered, the necessity for externalities and their implied simplistic, direct solutions becomes unnecessary. Coase (1990, p.27) emphasized: "It was unnecessary to use a concept such as 'externality' in the analysis to obtain the correct result." Externalities reflect irrationality, while transaction costs reflect rationality. Reintroducing transaction costs as a constraint can make seemingly irrational situations appear rational. Ignoring these real conditions and attempting forced corrections could result in even greater disasters than the original state. Admittedly, rationality does not imply absolute optimality and correctness; it denotes optimality and correctness within established conditions. However, only by understanding these conditions can we determine whether and how to make changes. Therefore, the practical significance of the two concepts is diametrically opposed.[20]

---

[19] Fart theory. Imagine two office workers A and B in an elevator. A farted and smelled to B. According to externality theory, A should compensate B. But the problem is, how much compensation should be paid? If the compensation amount is not fair, it might cause even greater harm. To accurately measure the compensation for A's fart, we would need at least the following elements: the compensation amount corresponding to the concentration of fecal odor per unit volume of air, a unit price for compensation, ideally regulated by law or a market fair price; and a gas collection device worn on the hip to measure the volume and concentration of gas. Now ask yourself, are you willing to wear such a device all day in order to accept and pay compensation? No one would, because it's better to endure 20 seconds than wear it for an entire day—and this is the logic of transaction costs. According to the logic of externalities, we should force others to compensate, but this creates high pricing costs and other "transaction costs." Instead of unwarranted meddling, it's better to leave it alone.

[20] Externalities, market failure, and welfare loss are concepts within the framework of general equilibrium theory.

Although "The Problem of Social Cost" has been later associated with the issue of externalities, Coase is actually calling for methodological innovation. Coase (1990, p.119) explicitly stated: "The aim of this article is to indicate what the economic approach to the problem should be." What Coase aimed to correct is precisely Arrow's perspective,[21] which immerses itself in externality thinking, viewing the world from an economic theoretical standpoint, and labeling all objective factors that make reality inconsistent with theory as "obstacles" and "losses," with economists tasked to eliminate these factors to align reality with theoretical requirements. Coase believed that the traditional approach of theory is akin to fitting a square peg into a round hole. The correct approach should start from the real world, identify various factors people consider when making choices, and reconstruct the economic analysis system using these factors. In other words, there is no absolute "optimal state" (the equilibrium point where supply and demand curves intersect) that transcends the real world, and we do not need concepts like welfare loss or market failure that come with general equilibrium theory. This is because, as long as the considered real conditions are sufficiently comprehensive, every situation is optimal. Therefore, transaction cost theory does not simply indicate the existence of various costs; it reflects a reconsideration of the analytical paradigm that includes only supply and demand factors, reasonably incorporating various real-world factors on which people base their choices. This is the practical significance of Coase's second perspective. Coase's discourse may be complex, but its meaning is straightforward: the real world is much more complex than economic models, hence, without investigation, there is no right to speak; decision-makers should refrain from making "off-the-cuff decisions," and economists should avoid offering "off-the-cuff suggestions."[22]

---

However, Coase's transaction costs constitute a challenge to general equilibrium theory, indicating they operate outside the same theoretical framework. According to Cheung's note, Coase even resisted the concept of "equilibrium" (see Cheung, 2019b, p.70).

[21] If you are fixated on the definitions of transaction costs and externalities, you will always be a second-rate economist. First-rate scholars focus on the underlying logic of the concepts, understanding how their creators and users perceive the world. If you don't approach these concepts from the perspective of their underlying thought "the way we think", but only focus on the wording of their definitions, you will never truly understand them. In the end, you'll be led astray by pseudo-scholars who, with muddled thinking, seek to deceive and mislead.

[22] If you want to know why the government introduces a policy, you should check historical records; if you want to know why young people are reluctant to have children, you should ask them directly. This is common sense, but

Arrow misinterpreted Coase severely. The subsequent scholars' heavy reliance on Arrow's definition amounts to nothing but a wild goose chase. Given Arrow's towering academic stature and influence, it's hard to imagine how detrimental and far-reaching the impact of this research path—focused solely on the form of transaction cost definitions while disregarding their nature—has been.[23]

Although both Coase's and Arrow's definitions have their shortcomings, in the early stages of theoretical construction, a flexible definition of basic concepts brought about tremendous expansion space for this emerging discipline. We must acknowledge that broad definitions like "costs using price mechanisms," "costs of market operation," and "costs of running the economic system" provided subsequent economists with the possibility to apply transaction cost theory to various aspects of social life such as firms, contracts, institutions, and even historical evolution, ultimately forming the expansive New Institutional Economics. However, as mentioned earlier, while Coase's and Arrow's definitions seem similar in literal expression, beyond the use of the term "transaction cost" in his discourse, Arrow's ideas have no connection to transaction cost theory. They were not discussing the same issue, and their theoretical perspectives were completely opposite. If subsequent economists could continue to explore the nature of transaction costs, transaction cost economics would not face the criticisms it does today. However, the promising start by Coase and others was not sustained, and after a short period of debate, transaction cost economics went astray.

## 2.2 Expansion of Transaction Costs

---

economists today are likely to engage in a highly technical but utterly useless task—running regression models, copying "golden quotes" from scholars in other fields, using obscure and hollow terminology to deceive themselves and their readers, making connections, and then publishing in an A+ journal. This is what our generation of readers sees—economists' works contain everything except economics, and the few ideas they do have are embarrassingly simplistic, yet these have come to define the leading thinkers of our time. This is why I often tell my friends, the closer you are to the "mainstream", the farther you are from real scholarship. Additionally, I have one more advice for those economists: economists are always eager to innovate in theory, always looking for things beyond common sense, but they always forget that the world is driven by common sense, and nothing in essence can transcend common sense.

[23] Even today, there are still people publishing papers on transaction costs every month! It's time for a clean-up! Let's continue the "Versuch einer Umwertung aller Werte"!

### 2.2.1 Concept Expansion: Critiques and Reconstruction by Alchian and Demsetz

Alchian and Demsetz's most renowned article on transaction costs and property rights theory is undoubtedly "Production, Information Costs, and Economic Organization." Alchian's "disciple" Cheung (2019a, p.32) succinctly summarized the core argument of this article: "It is an article on why economic organization or firms emerge, with the theme of shirk. It suggests that cooperation is beneficial to all, but delegated behavior requires supervision, leading to the emergence of firms with supervisory functions." Next, we will conduct a detailed analysis of this article.

In "The Nature of the Firm," Coase conceptualized the firm as a "employer-employee" relationship—due to the exorbitant costs associated with organizing production through the price mechanism, entrepreneurs opt to hire employees and utilize the power of administrative command to organize production. As a result, the firm emerges and partially replaces the market. Alchian and Demsetz questioned this assertion:

> "The firm does not own all its inputs. It has no power of fiat, no authority, no disciplinary action any different in the slightest degree from ordinary market contracting between any two people. I can 'punish' you only by withholding future business or by seeking redress in the courts for any failure to honor our exchange agreement. That is exactly all that any employer can do. He can fire or sue, just as I can fire my grocer by stopping purchases from him or sue him for delivering faulty products." (Alchian & Demsetz,1972, p.777)

As for the power of administrative command, it is simply the employer's continued involvement "in renegotiation of contracts on terms that must be acceptable to both parties" (Alchian & Demsetz,1972, p.777). The two authors pointed out that the "employer-employee" relationship is not the essence of firm organization. This is because administrative power remains within the scope of the contract, and neither the employee nor the employer can step beyond the agreed boundaries. Consequently, it can be inferred that there is no fundamental difference between firm organization and ordinary market contracts, leading to the conclusion that the assertion of "firms replacing markets" is incorrect.

The two authors did not stop there. They realized the disastrous consequences of shirking and delegation in the simple device of “team production”:

> “If his relaxation cannot be detected perfectly at zero cost, part of its effects will be borne by others in the team, thus making his realized cost of relaxation less than the true total cost to the team.” (Alchian & Demsetz,1972, p.780)

The key to solving this problem lies in supervision, which includes

> “measuring output performance, apportioning rewards, observing the input behavior of inputs as means of detecting or estimating their marginal productivity and giving assignments or instructions in what to do and how to do it. (It also includes, as we shall show later, authority to terminate or revise contracts.)” (Alchian & Demsetz,1972, p.782).

It is important to note that supervision is compensated, and the practical manifestation of these costs is the entrepreneur’s remuneration, that is, the entrepreneur’s residual claimant on employees (Alchian & Demsetz,1972, p.782). However, who should supervise the entrepreneurs? The authors’ answer was that, under market competition mechanisms, those who fail to fulfill their responsibilities will be eliminated, thus effectively constraining the behavior of entrepreneurs (Alchian & Demsetz,1972, p.781).

Thus, the two authors presented a theory of the firm that is diametrically opposed to Coase’s. Dietrich summarized the above view as follows:

> “Alchian and Demsetz (1972) have claimed that Coase’s analysis is, in a sense, a tautology. To suggest that firms exist because of the relative costs of using the market must be an empty truth; it would be just as valid to turn the framework on its head and suggest that markets exist because of the relative costs of management……The central problem here is lack of a rigorous theoretical framework.” (Dietrich,2008, p.14)

This critique is not without merit. If only the concept of “cost” is introduced without elucidating the nature of these costs, it becomes a theoretical sleight of hand, inevitably falling into Carnap's dilemma of "entelech".[24]

However, it is precisely here that the connotation of certain costs identified by Coase is affirmed, and the views of the two authors actually resonate with the first perspective introduced by Coase. Coase pointed out that without constraints, individuals incur significant costs in their market activities, and Alchian and Demsetz, in fact, indicated that the losses incurred due to shirking and delegation are the costs of lacking constraints, thus necessitating constraints on behavior. Demsetz later wrote that the 1972 article relied on "differences in shirking opportunities to understand the organization of the firm" (Demsetz,1988, p.151), but also noted that "it is true that transaction cost is involved in the existence of these problems, at least for a (too) broad definition of transaction cost." (Demsetz,1988, p.151), the two authors objectively propelled the development of transaction cost theory. Consequently, transaction costs are linked to the constraints on behavior.[25]

### 2.2.2 Competition: Alchian's New Perspective

Regarding behavioral constraints, Alchian went even further.[26] Alchian posed the following question:

"In every society, conflicts of interest among the members of that society must be resolved. The

[24] Biologist Driesch introduced the concept of "entelech" to explain biological activities and evolution, yet he failed to elucidate its meaning or provide new laws related to it. Therefore, Carnap concluded that this concept lacks explanatory power (see Carnap,1966, pp.14-16).

[25] In fact, the two authors hold views opposite to Coase's, as discussed by Cheung in the debate between "unpriced" and "discharged responsibility." Cheung reconciles this dispute by agreeing with Coase's perspective that "unpriced" (lack of constraints) leads to "discharged responsibility" (costs), rather than the reverse. However, what Coase did not point out is that "unpriced" is a result of choice (see Cheung, 2019a, pp. 208-217). This viewpoint is also reflected in the "Law of Substitution of Transaction Costs" introduced in the third part of this paper. While agreeing with Cheung's perspective, this debate does not affect the presentation of the relationship between transaction costs and behavioral constraints here.

[26] Alchian pointed out that competition is an essential process for determining the order of resource allocation, and his understanding of competition is deeper than that of anyone else in the world. Returning to the significance of the history of thought, we should not merely stop at citing Alchian's ideas but should think about how he brilliantly arrived at this insight—was it his unique life experiences? Or his distinctive perspective on the world? —think about how to think!

> process by which that resolution (not elimination!) occurs is known as competition. Since, by definition, there is no way to eliminate competition, the relevant question is, what kind of competition shall be used in the resolution of the conflicts of interest?" (Alchian,1965, p.816)

From there, he delved into competition, meticulously examining private and common property rights, and recognizing that different forms of property rights can induce behavioral changes (Alchian, 1965, p.828).

In another paper, Alchian provided a more systematic analysis of the relationship between different types of property rights and competition. Alchian (2006c, p.97) emphasized that private property rights "are extremely important in enabling greater realization of the gains from specialization in production," stating that they are "exist in principle but, quite sensibly, not be blindly and uncompromisingly enforced against all possible 'usurpers.'" (Alchian,2006c, p.103) At the same time, he (2006c, p.101) acknowledged that "not all resources are satisfactorily controlled by private property rights...... Other forms of control are then designed, for example, political or social group decisions and actions." However, this did not mean property rights were completely open:

> "If these other forms permit open, free entry with every user sharing equally and obtaining the average return, use will be excessive. Extra uses will be made with an increased realized total value that is less than the cost added; that is, the social product value is not maximized. This occurs because the marginal yield is less than the average to each user, to which each user responds....... Excessive costs will be incurred in competition for use of unowned, valuable resources." (Alchian,2006b, pp.101-102)

In other words, the constraints of various property forms were crucial in minimizing the decline in collective interest caused by excessive competition, even if private property rights were absent or ineffective, other property forms and institutional arrangements would fill this gap.

Subsequently, Alchian further refined the spectrum of competition:

> “Williamson: ‘opportunism’; Barzel: ‘measurement problem’; Wilson: ‘moral hazard’; Stiglitz: ‘asymmetric information with uncertainty’; Alchian and Demsetz: ‘shirking and monitoring’; Klein and Leffler, and Telser: ‘self-enforcing contracts’; Klein, Crawford, and Alchian: ‘vertical integration’; Strotz: ‘time inconsistency’; Goldberg: ‘right to be served’; Coase: ‘the firm’; Jensen and Meckling: ‘agencies.’” (Alchian,2006c, p.271)

All these concepts reflect issues of competition.

In summary, according to Alchian, competition is ubiquitous and inevitable, and the constraints on human behavior are essentially constraints on competition. If valuable assets are unowned, they will inevitably be overused, which highlights the necessity of property rights to avert this tragedy.[27] This extends Coase’s argument further: in Coase’s view, an unconstrained market operates at high costs, and people rely on property rights to reduce transaction costs and facilitate market transactions. Alchian added that unconstrained competition leads to a decline in collective welfare, and people use property rights to limit excessive competition. Ultimately, we find that the costs incurred by an unconstrained market and the decline in collective welfare are fundamentally the same issue.

### 2.2.3 Summary

Now, we can provide a preliminary summary. Coase identified a dual relationship between transaction costs and property rights, while Alchian and Demsetz introduced the perspective of constraining improper behaviors into the discussion of transaction costs. Moreover, Alchian pointed out that the essence of these improper behaviors is competition, which generates additional costs and thus must be constrained by property rights. We observe that transaction costs, property rights, and competition are interconnected, forming a theoretical framework of “Transaction Costs-Property Rights-Competition” that is emerging. The next steps involve further integration and refinement of this framework. Regrettably, this development has taken decades.

---

[27] My original wording was “the tragedy of the commons”. However, after reading a paper that distinguishes between the tragedy of the commons and the prisoner’s dilemma, I realized that my understanding was too superficial to make such a metaphor. Therefore, I decided not to use this metaphor to maintain rigor.

### 2.3 Went Astray: Form Over Substance[28]

#### 2.3.1 Review and Reflection

Transaction cost theory originated from two perspectives proposed by Coase, and ideally, it should conclude with accomplishing two corresponding tasks. However, as the theory has evolved, discussions have increasingly deviated from these tasks, becoming more formalized instead. There has been a rush to make the concept of transaction costs operational, focusing solely on how it is defined rather than discussing the nature of transaction costs. Consequently, transaction costs have been generalized into mere terms for specific costs like advertising fees, legal fees, and litigation costs. We must answer the fundamental questions that precede the introduction of any concept: what task do we intend to accomplish with this concept? Is this concept necessary and qualified to be used independently? When we use this concept, is there a new perspective underlying our thinking? History has not responded, but the echo resounds. To uncover the truth of history, we are going to make historical figures step down from the altar.

#### 2.3.2 Dahlman's Transaction Costs: A Repetition of Arrow's Defects[29]

Dahlman's definition can be summarized as follows: "The transaction process involves distinct, continuous stages, corresponding to three different types of transaction costs: search and information costs, bargaining and decision costs, and monitoring and enforcement costs." (Hu, 2019, p.30) This appears to be a standard and normative definition, yet Dahlman later elaborated in detail, revealing his understanding of the essence of transaction costs:

> "The task attempted in this paper is essentially twofold. First, although the role of transaction costs in the generation of externalities is well understood, no systematic analysis as yet exists of exactly what kinds of transaction costs are necessary to generate externalities. Thus, this paper will analyze the concept of transaction costs as it pertains specifically to externalities."

[28] The original title was "The Big Chaos".

[29] Dahlman represents the vast majority of people's understanding of transaction costs—transaction costs are externalities.

(Dahlman,1979, p.143)

He also claimed:

> "It is a very strange feature of modern welfare-policy prescriptions that they propose to do away with externalities, which are only one of the symptoms of an imperfect world, rather than with transaction costs, which are at the heart of the matter of what prevents Pareto optimal bliss from ruling sublime." (Dahlman,1979, p.161)

It is evident that Dahlman's problem closely mirrors that of Arrow. First, Dahlman believed that transaction costs generate externalities, but these two concepts do not belong to the same theoretical framework, and it is impossible to discuss a derivative relationship between them. Second, as we noted in our analysis of Coase's ideas, one major function of the transaction costs is to explain the rationality behind people's tolerance of so-called "negative impacts". Thus, transaction costs are not a hindrance to achieving Pareto optimality; rather, they are crucial in facilitating it (the status quo). For this reason, despite Dalman's definition has received much attention, it lacks practical value.

### 2.3.3 Williamson's Transaction Costs[30]

Williamson's theory of transaction costs is founded on three key elements: bounded rationality, opportunism, and asset specificity (Williamson,1985, p.42). He (1985, p.41) also posited that transactions are the fundamental unit of economic analysis, and any problem can be reformulated as a contractual issue. He (1985, p.20) distinguished transaction costs into those "ex ante and ex post types". Ex ante transaction costs refer to "the costs of drafting, negotiating, and safeguarding an agreement" (Williamson,1985, p.20). Ex post transaction costs include costs of maladaptation, haggling, setup and running, and effecting secure commitments (Williamson,1985, p.21). In summary, Williamson defined transaction costs as "costs associated with contracting." However, as observed, Williamson's definition does not explain why human behavior needs to be constrained, nor does it integrate various real-world factors into economic analysis. It merely emphasizes that exchange and contracting entail costs (see Williamson,1985, p.20), falling short of fulfilling Coase's two tasks.

[30] When criticizing Williamson's views, I felt a deep sense of helplessness—since Williamson said nothing, I didn't even know where to begin. Fortunately, Demsetz articulated this point for me, or else this section of my paper would have remained blank.

Moreover, as a staunch defender of the theory of transaction costs, Williamson himself did not actively employ this concept. Williamson claimed:

> "The inordinate weight that I assign to transaction cost economizing is a device by which to redress a condition of previous neglect and undervaluation. An accurate assessment of the economic institutions of capitalism cannot, in my judgment, be reached if the central importance of transaction cost economizing is denied." (Williamson,1985, p.17)

However, it is rather awkward that, as Demsetz(1988, p.147) noted, "Williamson uses the first part of his book about the institutions of capitalism to claim that its foundation lies in transaction cost considerations, but he fails to make substantive use of transaction cost throughout the remainder of the book". This raises a question: As successors, how can we continue to trust the effectiveness and practicality of the Williamson framework when even Williamson himself did not actively use the transaction costs he defined?

### 2.3.4 Dietrich's Transaction Costs

The focus of Dietrich's work, "TRANSACTION COST ECONOMICS AND BEYOND—Towards a new economics of the firm," is on the firm rather than the concept of transaction costs. However, Dietrich (2008, p.14) believed that developing an appropriate theory of the firm requires "moving outside a narrow transaction cost perspective".

Dietrich did not fully agree with Williamson's views. He (2008, p.24) argued that the introduction of "bounded rationality" has caused confusion—Williamson's transaction cost is proposed within the framework of bounded rationality, while production cost is proposed under the framework of completely rationality, these two concepts "cannot be just thrown together." Furthermore, Dietrich (2008, p.25) also pointed out that Williamson failed to properly recognize the relationships between bounded rationality, uncertainty, and opportunism, which has led to the central role of opportunism being to introduce "logical problems into transaction cost economics."

Dietrich's argument holds merit, yet it is curious that although he undermines two of the three pillars of Williamson's framework, his own definition differs from Williamson's only in its specific phrasing, not in content. It still falls within the "ex ante" and "ex post" categories and fails to fulfill the two tasks of transaction cost theory. Dietrich (2008, p.28) stated: "we can define transaction costs in terms of three factors: search and information costs, bargaining and decision costs, and policing and enforcement costs."

### 2.3.5 North's Transaction Costs: The Anti-Transaction Costs Masquerading as Transaction Costs

North (1990, p.27) anchored transaction costs in "the costliness of information" and further decomposes transaction costs into measurement costs and enforcement costs (1990, p.32). Correspondingly, North (1990, p.28) also defined transformation costs: "the resource inputs of land, labor, and capital involved both in transforming the physical attributes of a good (size, weight, color, location, chemical composition, and so forth)." Consequently, he (1990, p.28) outlines the relationship between production costs and transaction costs: "the costs of production are the sum of transformation and transaction costs." North's approach is essentially common sense—he replaced the everyday concept of "total cost" with "production cost," defining the manufacturing-related portion as transformation costs, which closely aligned with the traditional notion of production costs, and the information-related portion as transaction costs. In other words, North merely disaggregates the price components of goods.

In practical application of the concept of transaction costs, let us examine an example provided by North:

> "The transaction costs of the transfer are partly market costs - such as legal fees, realtor fees, title insurance, and credit rating searches - and partly the costs of time each party must devote to gathering information, to searching, and so forth. Obtaining information about crime rates, police protection, and security systems entails search costs to the buyer. To the degree that the buyer's utility function is adversely affected by noisy neighbors or pets, it will pay to invest in ascertaining neighborhood characteristics and the norms and conventions that shape neighborhood interactions." (North,1990, pp.62-63)

In just the matter of house transfers, North introduced 11 types of costs, illustrating his effort to capture the various detailed costs individuals face.

In their seminal article on measuring transaction services in the U.S., written with Wallis,[31] the authors interpreted the transaction costs they attempt to measure as the actual costs incurred:

> "Our notion of transaction services and transaction costs is perfectly analogous to the notion of market income and total income in the national income accounts. GNP does not claim to measure the total income of individuals in a society, but the income that individuals generate through the market process (aside from imputed nonmarket items, such as owner-occupied housing and nonmarketed farm output). In the same way transaction services capture only that part of transaction costs that flows through the market." (Wallis & North,1986, p.99)

They further emphasize in the same article (1986, p.121) that "the costs of transacting may have been as much a limiting factor on economic growth as transformation costs……Until economic organizations developed to lower the costs of exchange we could not reap the advantage of ever greater specialization." This reveals that the authors viewed transaction costs within the framework of the "specialization and exchange relationships," where higher levels of specialization result in higher costs of exchange. In addition, according to their calculations, transaction costs that flow through the market in the U.S. in 1970 exceeded 40% of that year's GNP (Wallis & North, 1986, p.121).

North's transaction costs have three shortcomings. First, the view that transaction costs (and transformation costs) are impediments to economic growth lacks substantive meaning, as we can reasonably argue that any cost is a hindrance to economic growth—this merely reiterates the definition of "cost" without demonstrating any unique role of transaction costs.[32] Second, North does not fulfill Coase's aim of using transaction costs as a starting point to revolutionize traditional economic theory (i.e., the two perspectives and two tasks). On the contrary, North opposed the idea of transaction costs as a brand-new theory:

> "From the viewpoint of the individual both of these functions are 'productive'; that is, transaction and transformation costs are incurred only if the expected benefits from doing so exceed the costs of doing so. The behavioral similarity of transaction costs and transformation costs is critical, since it implies that we do not need a new 'transaction costs theory' of human behavior to deal with transaction costs; simple price theory will suffice." (North,1986, p.97)

[31] It should be noted that the earlier citation of "Institutions, Institutional Change and Economic Performance" "restate" the discussion on transaction costs and transformation costs in this paper (North,1990, p.28).

In other words, North does not recognize "transaction costs" as an independent concept or a proprietary theory; essentially, he views them as an extension of production costs, as also reflected in the previous context. Thus, North was not a constructor of transaction cost theory at the time but rather a critic.[33] Third, the Walrasian system has been criticized by transaction costs theorists for ignoring transaction costs, including North, and transaction costs have no place in North's depiction of the Walrasian system (see North, 1990, p.30). However, understanding transaction costs from the perspective of GNP, which measures transaction costs based on the market turnover of the transaction sector, makes the transaction cost theory compatible with the Walrasian system. The Walrasian system encompasses numerous industries, including those transaction sector; thus, it does not overlook the transaction costs on a macro level, leading to a contradiction between North's arguments and his method of calculation.

Evidently, North's issue essentially stems from neglecting to explore the nature of transaction costs,

---

[32] North's understanding of the social impact of transaction costs reflects the perception of most people. However, the biggest problem with this understanding is that if you remove the word "transaction" from "transaction costs", the conclusion of the analysis remains unchanged.

[33] I have not seen any scholar mention this.

leading to a series of cascading problems—due to the failure to discern the nature of transaction costs, they are conflated with production costs; and because of the inability to distinguish these two types of costs effectively, transaction cost theory is unwittingly assimilated by the very systems it intends to critique.

### 2.3.6 Barzel's Transaction Costs

Barzel's definition, which directly links transaction costs to property rights, has been widely accepted: "I define transaction costs as the costs associated with the transfer, capture, and protection of rights." (Barzel,1997, p.4) If we consider only the literal meaning of this statement, Barzel seemed to emphasize the connection between transaction costs and property rights. However, making a judgment based solely on this statement can also be seen as taking it out of context or interpreting it superficially. Importantly, Barzel (1997, p.4) expounded in a footnote to this definition that his understanding of transaction costs is equivalent to the agency costs defined by Jensen and Meckling, which arise from the misalignment of interests between principals and agents, including "the monitoring expenditures by the principal," "the bonding expenditures by the agent," and "the residual loss" (Jensen & Meckling, 2019, p.164). Clearly, the literal meaning of Barzel's definition has limited connection to the thoughts of these two authors, and even Barzel himself (1997, p.13) points out in his text that the principal-agent model is not a property rights model. Therefore, to uncover the true meaning of Barzel's definition, we need to consult additional discourse for reference.

Dramatically, in the very next sentence, [34]Barzel (1997, p.4) shifted the focus to information:

> "If it is assumed that for any asset each of these costs is rising, and that both the full protection and the full transfer of rights are prohibitively costly, then it follows that rights are never complete, because people will never find it worthwhile to gain the entire potential of 'their' assets. In order that the rights to an asset be complete or perfectly delineated, both its owner and other individuals potentially interested in the asset must possess full knowledge of all its valued attributes. With full knowledge, the transfer of rights to an asset can be readily effected. Conversely, when rights are perfectly delineated, product information must be costless to obtain and the (relevant) costs of transacting must then be zero."

[34] I guess the reason people failed to uncover Barzel's true intent is simply because he used endnotes in this book, and scholars were too lazy to turn the pages, resulting no one read the annotation. This is already foolish enough, but even more foolish is that people overlooked the sentence that follows it.

This indicates that property rights and information costs are not parallel; the costs associated with property rights are essentially expenditures on information. Later, Barzel (1997, p.53) explicitly stated, "information problems are at the heart of the high cost of transacting." From this, we can conclude that the thread linking Barzel's various points is not property rights, but information costs.

Barzel's views on information costs also require revisiting his paper published in 1985. Barzel noted that the existence of transaction costs adds complexity to competitive behavior, and fundamentally, transaction costs are information costs. If information costs were zero, there would be no transaction costs, and zero information costs would imply omniscience (see 1985, pp.6-7). Combining all the above, we can conclude: To avoid ambiguity, Barzel's definition could and should be framed in terms of "costs of information."

Barzel's theory has two shortcomings. First, Barzel constructed a framework of "Information Costs → Transaction Costs → Economic Phenomena", but in this framework, it is actually the information costs that are explanatory, not the transaction costs. Substituting transaction costs with information costs has no detrimental effect.[35] According to Occam's Razor, since the essence of the problem (information costs) has been identified, there is no need to retain the non-explanatory intermediate form (transaction costs), and thus the concept of transaction costs is unnecessary in this system. Second, Barzel only considered the two extreme cases where information costs are zero and information is completely lacking. In real life, non-zero information costs represent a significantly complex situation, and analyzing only the “infinite” and “none” scenarios is clearly insufficient. Regrettably, the discussion has not been continued. As Wang (Barzel, 2017, p.3) mentioned in the preface to the Chinese translation of “The Economic Analysis of Property Rights”, “In Barzel’s theory of property rights, the concept of ‘transaction costs’ no longer plays a primary role.”

[35] In the third edition of "The Economic Analysis of Property Rights", Barzel removed this note and corresponding chapter, replacing it with a critique of the confusion between transaction costs and information costs. This indicates that Barzel later recognized the inherent flaw in his "information costs → transaction costs → economic phenomena" framework. Unfortunately, since he did not address the fundamental issue of the cost of achieving self-maximization, Barzel's argument ultimately did not undergo any substantial change.

### 2.4 Reorganization and Analysis

A distinctive feature of new institutional economists is that each one forms their own thought, with individual emphases, thus the evolution of their ideas is not strictly linear. The considerations of the scholars mentioned above essentially cover all the key points of debate concerning the concept of transaction costs. I will organize their thoughts into several propositions for analysis.

#### 2.4.1 Transaction Costs as an “Impediment”

The following assertion is quite common: if transaction costs (or their other representations, such as information costs) were zero, the resources currently invested in acquiring information, negotiating, and bargaining could be used for productive activities, thus naturally leading to the conclusion that “reducing transaction costs enhances economic performance.” However, viewing transaction costs as an impediment to transactions is as absurd as viewing production costs as an impediment to production, because this view does not base on maximizing within constraints but rather from lamenting that these constraints prevent achieving an illusioned maximization. Everyone desires lower costs, and this argument simply expresses a hopeful wish to reduce all costs. Following this logic, one could also argue: if sleep did not require time, people could use the time normally spent sleeping for productive activities, thus sleep impedes economic growth. Apparently, this is a meaningless assertion.

People have a variety of choices, each with its own set of costs. Economists take the costs associated with the choices people actually make—the costs of the choices that maximize their interests—and label part of these costs as transaction costs, recognizing them as a decline in economic performance. This is somewhat peculiar. On one hand, we believe that incurring transaction costs is efficient (maximization for the individual); on the other hand, we regard incurring transaction costs as inefficient (decrease economic performance), creating a logical contradiction. So, what do the terms 'inefficiency' and 'hindrance' refer to? Apparently, a new theory is needed.

### 2.4.2 Transaction Cost "Detailism"

Reducing various minutiae of costs directly back to economic analysis is likewise a fallacy. The problem with general equilibrium theory is evident, there is no place for currency and institutions, resulting in significant discrepancies with the real world. However, the theoretical errors of general equilibrium should not obscure its correctness of its thought, namely simplifying the real world. Real world factors are infinite, even within the realm of exchanges, rendering economics a simplistic and verbose description of human choice if all conditions were directly incorporated into analysis, merely swinging from one extreme to another.

The goal of economic theory is to identify key factors and simplify the world. If every nuanced factor must be considered for analysis and causal inference, economics loses its value, and the nature of transaction costs becomes ambiguous. Thus, while transaction costs may comprise an infinite array of intricate expenses, yet the analysis of transaction costs cannot be so exhaustive. There evidently lacks a crucial intermediary concept between the two.

### 2.4.3 Transaction Costs and Production Costs

As Carnap (1966, p.15) pointed out: "It is not sufficient, for purposes of explanation, simply to introduce a new agent by giving it a new name. You must also give laws." Applied to the issues discussed in this paper, Carnap's assertion implies that simply invoking the concept of "transaction costs" without specifying its meaning lacks explanatory power in addressing economic problems. Therefore, if we aim to use transaction costs to explain economic phenomena, we must clarify the nature of transaction costs and the fundamental distinction between these costs and production costs.

Costs inherently imply the maximization of self-benefit, and in practical usage, "production costs" are often abstractly understood as "costs." Hence, production costs are effectively used to express the maximization of self-benefit, as evident in North's discourse. However, as mentioned earlier, any production costs or transaction costs incurred by individuals serve the same purpose of self-maximization. From this perspective, the transaction costs and production costs discussed by these scholars are fundamentally indistinguishable. This underlies the difficulty in economically differentiating between the two types of costs. Thus, the debate between transaction costs and production costs shifts from an economic problem to an accounting classification issue—essentially, the debate revolves around whether advertising expenses should be recorded as independent transaction costs or included in production costs. As we can see, within the existing theoretical framework, if we desire independence for the concept of transaction costs, it necessitates a reduction in the scope of production costs, providing space for transaction costs to exist independently.

However, from this perspective, transaction costs have not altered the awkward position where firms are still understood as production functions; they merely make the production function flatter. They also fail to achieve the innovation in the general equilibrium method desired by Coase; they merely shift both the supply and demand curves to the left, a feat achievable by any type of cost. Therefore, if transaction costs are to be regarded as an independent concept, they cannot fundamentally represent the cost of achieving self-maximization. Unfortunately, both "information costs" and "exchange costs," as defined by the scholars mentioned above, contradict this requirement.

### 2.4.4 The Value of Transaction Costs

Before introducing a concept into economics, we must clarify what problem we intend to address with it. The value of the transaction cost concept lies in Coase's two perspectives and their corresponding tasks. The correct approach should be to return to Coase and fulfill these two major tasks. Unfortunately, the scholars mentioned above have not accomplished this.[36]

Previous research seems to imply that the value of the transaction cost concept stems from the fact that "transactions involve costs." However, a more general fact is that all human actions incur costs.

[36] People were wrong because they didn't think in the way Coase thought.

Choosing to transfer property incurs a cost, as does choosing not to transfer property; gathering information incurs a cost, as does choosing not to gather information. This is common knowledge, understood by people without the need for economists. Apparently, the operation of price systems incurs costs, designing and signing contracts incurs costs, and monitoring individuals incurs costs. Given this, why do we spend transaction costs and bother ourselves? Clearly, the key issue is not merely listing the specific costs incurred by individuals, but exploring the economic implications of these costs—why individuals are willing to incur them and the impact of such behavior on society. Few scholars who have addressed this issue either mistakenly label transaction costs as "obstacles" or simply repeat the presumption of "self-maximization." However, this does not introduce a new perspective into economic theory and thus contradicts the original intention of transaction cost economics. We can boldly imagine whether economists, when fervently writing sentences like "Coase taught us that market transactions involve costs," realize the absurdity of their words? This interpretation of transaction costs undoubtedly oversimplifies and diminishes Coase's ideas and shows excessive disdain for the intellectual capabilities of economists before Coase's fame. North's contradictory discourse precisely demonstrates that those criticized scholars were not unaware that transactions incur costs; instead, they regarded them as part of existing theoretical models. They just did not specifically emphasize this point, much like subsequent scholars do not particularly emphasize that "blinking incurs costs."

The concept of transaction costs does not inherently possess economic significance simply because it is "overlooked by economists." In fact, in the process of simplifying the world, countless costs are not included in analysis. If overlooking transaction costs requires, using Williamson's words, "redressing a condition of previous neglect and undervaluation," then we should also consider the "cost of breathing," "cost of blinking," and various other costs, advocating for minimizing them as much as possible. After all, these costs are similarly omitted from traditional economic analysis. While we could incorporate such costs into transaction costs, the outcome is evident—transaction costs would lose any distinctive meaning, becoming indistinguishable from general costs, rendering them unnecessary. There is evidently a question of "degree" here, but the scholars mentioned above did not engage in this discussion.

### 2.4.5 Reflection and Prospects[37]

A notable phenomenon is that, despite the absence of a recognized, explicit definition of transaction costs in academia, the various expressions used do not fundamentally differ in their practical implications. The most typical example is Dietrich's criticism of Williamson's ideas. This phenomenon undoubtedly illustrates that economists have reached a certain consensus on the concept of transaction costs, yet their definitions bear little connection to the theoretical foundations they painstakingly construct.[38] As for the sluggish development of New Institutional Economics today (it may even be a regression from its original intention of innovating traditional theories), the ambiguity in the concept of transaction costs bear considerable responsibility. However, as we can see, the lack of a unified formulation is merely a superficial manifestation of the ambiguity surrounding transaction costs. The root of the problem lies in the theoretical deficiency itself—few have attempted to explore the nature of transaction costs, instead, adjustments are made to expressions, as long as a definition roughly aligns with people's imagination, it is considered a definition of transaction costs. This is also why Arrow's concept of transaction costs has gained such widespread acceptance.

---

[37] There are three reasons why people get it wrong from start to end. First, many scholars discuss transaction costs not from Coase's way but through their own interpretation. However, this interpretation is essentially just imposing inherent biases from different dimensions onto transaction costs, and will only produce the same level of garbage. And those scholars who over interpret it are disgusting—transaction costs are costs, knowing what they are (qualitative analysis) and how to measure them (quantitative analysis) is enough, why some even drag Kant into the discussion? This behavior, if put nicely, is called offering new perspectives; more realistically, it's academic enclosure—in order to become a pioneer in a certain field, they use the power of their early career or seniority to fill it with garbage, so that even if future thinkers have better ideas, they'll still be called predecessors. Second, many claim to have read certain books or articles and cite them in their papers, but they likely haven't read them, which explains the frequent misquotations and misinterpretations in the study of transaction cost economics. As for their citations, they might be unmarked secondary quotations or simply retrieved through keyword searches. Third, most scholars prefer to use indirect citations, which don't require page numbers, creating a vast space for unscrupulous scholars to fabricate. That's why I use direct quotations whenever possible.

[38] The fact is, the so-called "foundation of the theory" that scholars refer to serve only to make the author appear to be professional and their thinking rigorous, but there is no help to the systematization or rigor of the theory. This is another "emperor's new clothes", but for some reason, later scholars are particularly eager to discuss these non-existent clothes. This once again proves that if you don't consider the author's motivation for writing this passage, their underlying thought process, and "the way they think", you will never figure out the main point.

If we continue down this path, the end of transaction cost theory would be economists listing one cost after another to enrich the economic dictionary. However, as we have argued, merely pointing out that something incurs a cost contributes nothing to the advancement of transaction cost theory—this is precisely why almost all scholars' theories on transaction costs ultimately end up fruitless.[39] Up to this point, there is no definition of transaction cost that is sufficient to distinguish transaction cost from production cost. Whether this concept should even exist remains debatable, and it certainly fails to accomplish the task of revolutionizing the economic paradigm. The term "transaction costs" simply serves as a generic term for a range of costs such as "advertising fees," "legal fees," "litigation costs," and so on.[40] Even if this concept holds economic significance, it does not originate from the concept of "transaction costs" itself, but rather from the characteristics of those specific costs. So, what are the characteristics of these costs? People's answer is: hindrance and the costs of self-maximization.

At this moment, the theories of firm theory, contracts theory, and institutions theory, which are built around traditional transaction cost theory and have long dominated economics textbooks, appear rather ludicrous. The existence of these theories can only signify one thing—despite the confident assertions of transaction cost economists, but they have no idea what they are talking about.[41]

---

[39] It should be noted that I do not imply the previous achievements are worthless, but rather unsuitable as direct definitions of transaction costs. In the fourth section of this paper, I will refer to the frameworks of other scholars to indicate that the transaction costs as perceived by the economists are simplifications of some fundamental costs.

[40] For example, in finance economics, people refer to securities transaction commissions as transaction costs. I really can't understand, what practical significance renaming fees as transaction costs has for the development of financial theory? Why can't I call securities transaction commissions what they are, but have to call them "transaction costs"? The following statement is purely common sense—calling a tomato a potato doesn't make it a mango. Economists have no common sense while they have a good command of advanced words, do readers find this strange? Let's return to common sense: If someone doesn't recognize mathematical symbols but can write a full board of mathematical formulas, then their writing is either copied or nonsensical.

[41] Despite the language of these economists are full of logical confusion, and they have no idea what they are talking about, however, their vague definitions of transaction costs serve a clear purpose. Now, let's revisit the history of the transaction cost concept: Coase introduced transaction costs to challenge the effectiveness of a purely free market; in "The Problem of Social Cost", Coase expanded the idea to oppose the logic of externalities and the "visible hand" of government; Lees and Rice introduced transaction costs to criticize Arrow's overconfidence in government enforcements; Arrow introduced transaction costs to justify the role of government

## 2.5 Commons: Transaction rather than Exchange[42]

The study of property rights by the New Institutional Economics (NIE) evokes memories of a school of thought active in the late 19th and early 20th centuries, keenly focused on ownership issues—the American Institutional School. For ease of distinction, later scholars classified them into the Old Institutional School and the New Institutional School, according to their chronological appearance. Although the two schools share overlapping areas of focus, their viewpoints and methodologies are fundamentally incompatible. Coase (1998, p.72) once criticized the research of the Old Institutional School as nothing more than a collection of "without a theory to bind together their collection of fact." Rutherford summarized this phenomenon, stating

> "Members of the NIE do not see their work as a continuation of the endeavours of old institutionalists such as Veblen or Commons, but as a distinct effort to apply more standard economic approaches to institutional issues. Conversely, old institutionalists have been unimpressed by the NIE, regarding it as a part of a research tradition they rejected long ago." (Rutherford,1996, p.181)

However, this divergence does not overshadow the foresight of key figures in the Old Institutional School. Commons, for instance, regained the attention of economists decades later due to his insightful intervention; later scholars introduced transaction costs, some to revive the myth of the free market, others to revive the myth of a strong government—however, all of these are just myths. The concept of transaction costs, originally a ticket from Coase to help people back to reality, ultimately became a worthless piece of paper for rewriting myths. Since the concept of transaction costs was discovered, it has been used as a tool for scholars to support either the government or the market. Except for its creator, no one can think about transaction costs without bias, nor does anyone really have the interest to figure out what transaction costs truly are. When people want to emphasize market forces, transaction cost theory supports the market; when people want to emphasize government power, transaction cost theory supports the government. Everything is simply a transaction, and those renowned papers are just invoices that scholars write for politicians or other interest group (I'm not a left)—now do you understand why transaction cost economics is so logically chaotic yet goes unnoticed? Now do you understand why I am in great angry?

[42] I once considered using "Back to Commons" as the theoretical foundation for my paper, but I realized there was no need to drag a passed figure in to support my own academic work. I also had plenty of room to quote some philosophical maxims to boost the "value" of this paper, but I always recall a warning I once gave myself: "Answering an economic question with a philosophical proposition only shows the economist's lazy".

understanding of “transactions”.

### 2.5.1 Commons’s Distinctive Research Approach

Commons, a seasoned legal professional with experience as a laborer, visits to businesses, formal training in economics, involvement in state legislation, and later forays into politics, possessed genuine and vivid observations and experiences across various aspects of social life. Commons’s unique background endowed him with a distinct perspective on observing economic and social phenomena—he examined case law as his research subject, understanding the world from the standpoint of legal judgments and rights (Commons,1990, p.3). The former angle corresponds to his studies on transactions, while the latter corresponds to his research on property rights, both of which converge in legal practice. It is worth noting that his conceptualization of "transactions" as the basic unit of economic analysis did not stem from an examination of exchange processes in social reproduction but rather from the notion of “integrating legal institutions into economics” (Commons,1990, p.3). In fact, this is precisely what Coase vocally advocated for.

### 2.5.2 Commons’s Interpretation of Transaction

Commons (1990, p.6) drew from Hume and Malthus’s concept of “scarcity” as the theoretical foundation. He recognized that the scarcity of resources dictates that conflicts of interest among individuals are commonplace, contrary to the classical assumption of abundance. Commons astutely observed that rectifying the errors of classical economics necessitates identifying a marker reflecting interpersonal, rather than human-object relationships—ownership. He stated:

> “The courts of law deal with human activity in its relation, not of man to nature, but to the ownership of nature by man. But they deal with this activity only at a certain point, the point of conflict of interests between plaintiff and defendant. But classical economic theory, based on relations of man to nature, had no conflict of interests in its units of investigation, since its units were commodities and individuals with ownership omitted. These ultimate units produced, in fact, along with the analogy of equilibrium, a harmony of interests rather than a conflict of interests. Hence the ultimate unit to be sought in the problem of correlating law, economics, and ethics is a unit of conflicting interests of ownership." (Commons,1990, p.57)

Concurrently, Commons (Commons,1990, p.57) acknowledged that from scarcity arises not only conflict but also "interdependence", which when established long-term, transforms into "order".

Thus, Commons, through relentless examination of Supreme Court precedents and cases of collective bargaining, labor arbitration, and commercial arbitration, synthesized the interactive model of "conflict-dependence-order." It is within this framework that Commons (1990, pp.57-58) identified his most suitable fundamental unit—transactions. In Commons's framework, transactions differ fundamentally from exchanges: exchanges involve the actual transfer of goods, while transactions involve a transfer in legal rights (Commons, 1990, p.58). The crux here lies in the involvement of ownership, marking the relationship between individuals.

It is worth noting that Commons did not introduce the concept of transaction costs, nor did he (1990, p.115; p.123; p.176; p.254; p.378; p.380) question the Walrasian system like modern scholars. Considering that Commons is an economist who particularly emphasizes the importance of "transactions", we cannot simply regard this as Commons' carelessness.

### 2.5.3 Summary

In summary, scholars of the New Institutional Economics tend to focus primarily on the exchange phase within social reproduction, resulting in their transaction costs often equating to "exchange costs" (the actual costs incurred in transactions). Conversely, Commons, as an early explorer, was not constrained by the economic paradigm. Commons's theoretical contribution lies in successfully extracting the essence of transactions from the interactive model of "conflict-dependence-order," expanding the concept beyond mere market transactions to various aspects of social life. From this perspective, Commons's conflict theory shares similarities with Alchian's theory of competition.

Of course, there are certain differences between the two as well. Regarding the process of conflict or competition, Commons treated the transition from conflict to order as natural, whereas Alchian emphasized the associated costs. In terms of the outcomes of conflict or competition, Commons attempted to discuss general results, suggesting that conflict ultimately leads to order, which is collective control over individuals, essentially institutionalization (Commons, 1990, p.6; p69; p.73); Alchian, on the other hand, provided specific solutions, suggesting that unrestrained competition leads to chaos, necessitating property rights as a constraint to resolve this situation. Regarding ownership or property rights, limited by the legal practices and theoretical level of his time, Commons did not clearly distinguish between ownership and property rights. Despite differences in specific terminology, we find that for Commons, ownership reflects interpersonal relationships, while for Alchian, property rights serve as constraints on competition.

Compared to Commons, Alchian's ideas were clearer and more explicit. However, we must acknowledge that Commons, as a figure from over a century ago, had ideas that transcended his era. Nonetheless, it is regrettable that Commons did not organize his groundbreaking ideas into a concise and effective economic methodology. As a result, his viewpoints remain inspirational rather than constructive. This is the fundamental reason why his ideas have been long forgotten.

### 3.The End of the Road: Cheung's "Trinity"

Cheung is one of the founders of New Institutional Economics. The neglect of Cheung's ideas is truly baffling, particularly in the reception of his work "Economic Explanation." Cheung regarded "Economic Explaination" as his finest and most significant economic work in his lifetime. (Cheung, 2019b, pp.19-20). As early as the first volume, Cheung expressed dissatisfaction with the current state of New Institutional Economics, eagerly announced his intention to "clean up" New Institutional Economics in the fourth and fifth volumes (2019b, p.205). However, it is perplexing that despite the entire fourth volume revolving around transaction cost theory, few scholars have paid attention to his in-depth exploration. Mention of Cheung's definition of transaction costs as "costs do not exist in a Robinson Crusoe economy" and "institution costs" represents only a small portion of his contributions, with the truly critical aspects remaining largely undiscussed.

Cheung combined the ideas of Coase and Alchian (Cheung, 2019a, p.40), achieving a singular accomplishment by establishing a theoretical framework of "transaction costs-competition-property rights," which predecessors had not accomplished. We can summarize his approach as a "trinity." This "trinity" refers to the unity of transaction costs, rent dissipation, and institution costs, each excelling in explaining different types of issues. Chueng stated:

> "Transaction costs, institution costs, and rent dissipation are three different perspectives on social costs...Generally speaking, examining market contracts from the perspective of transaction costs is most appropriate because it is the most direct. Examining issues outside the market, such as customs, religion, and status hierarchy, from the perspective of institution costs is more suitable because these are general constraints rarely involving bargaining behavior in the market. And rent dissipation? It is most applicable in explaining competitive behavior, as both market and non-market competition can easily be viewed from the perspective of rent dissipation." (Cheung,2019a, p.84)

Next, we will delve deeper into each component.

## 3.1 Transaction Cost

The connotation of transaction costs is closely linked to the underlying economic methodology, and different approaches lead to varying understandings of the nature of transaction costs. In general equilibrium analysis, every aspect of the real world is abstracted into countless supply and demand curves. Economists analyze from both supply and demand sides, treating transaction costs as factors outside of supply and demand, acting as frictional "forces".

Coase's second perspective suggests that general equilibrium theory is overly simplified, and economists should reconstruct analytical frameworks based on people's real conditions. However, Coase's expression is not entirely clear; he only roughly points out the direction without providing a clear methodology.

Cheung, at Coase's behest (Cheung, 2019c, p.284), pursued Coase's path to the extreme, opting for the path of "economic explanation," which entails "using the perspective of economics and scientific methods to explain phenomena or human behavior" (Cheung,2019b, p.42). Specifically, Cheung discarded the general equilibrium approach that abstracts the economic society into only supply and demand factors. Instead, he built a theoretical framework based on "maximizing behavior under constraints," comprising only the principles of demand, cost, and competition, leaving the remaining space for transaction costs, which he termed "constraints" (Cheung,2019b, pp.89-90). The term "constraints" refers to additional circumstances in theoretical analysis (Cheung, 2019b, p. 69). Cheung emphasized the importance of observing the real world, stating that explanations lacking empirical evidence would be deemed devoid of economic content. Thus, the responsibility of providing empirical evidence fell upon transaction costs, thereby achieving the methodological innovation Coase had not achieved.

Thus, Cheung reached the culmination of Coase's second perspective (task), where he identified the essence of transaction costs—constraints—that Coase had intended but failed to pinpoint successfully. This process did not evolve solely from continuous exploration of the concept of transaction costs, but rather from a thorough transformation of traditional methods. However, Cheung's assertion (2005, p.104) that "at least 80% of GDP comes from transaction costs" indicates that transaction costs here are still "part of the costs of maximizing oneself" in practical usage, similarly facing the dilemma of North's transaction costs.

### 3.2 Rent Dissipation

Cheung's concept of rent dissipation stems from Alchian's oral tradition regarding property rights and competition (Cheung, 2019a, p.29). It implies that "valuable resources, due to competition among individuals, partially or completely lose their value" (Cheung,2019a, p.79). In his analysis of open-sea fishing, Cheung illustrated the concept of rent dissipation (see Figure 1).

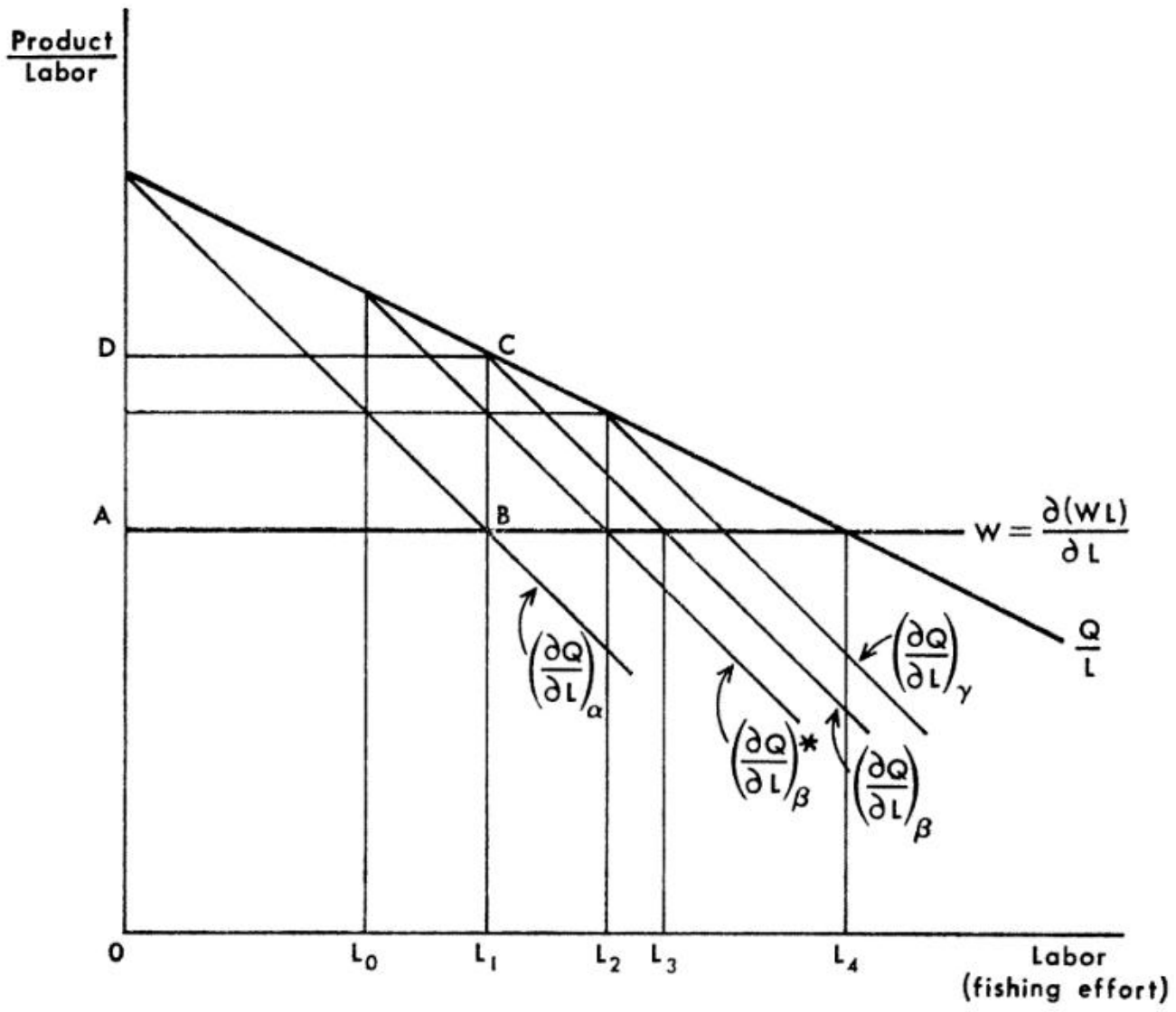

**Fig. 1** Rent dissipation (Cheung,2005, p.185)[43]

Apparently, the optimal utilization of resources occurs at the intersection of the marginal product curve and the marginal cost curve (which corresponds to the average cost curve in Figure 1). The quadrilateral ABCD above the average cost curve represents rent, which is the surplus income resulting from the optimal utilization of resources. However, the consistency between labor input and the required input at the optimal point is ensured by the restriction on total labor input by resource owners under private property conditions. In the absence of this constraint, unrestricted competition leads to an equilibrium point at the intersection of the average product curve and the marginal cost curve (the average cost curve), rather than the optimal point of resource utilization. With each additional competitor, the surplus income (rent) of the asset decreases, and when countless competitors enter the market, rent decreases to zero, turning high-quality assets into mediocre ones (see Cheung,2005, pp.184-190).[44]

---

[43] Influenced by the queueing model, for a long time, I misunderstood the concept of rent dissipation (Barzel was right; the misunderstanding was due to my own lack of deeper research). It wasn't until one day that I realized rent dissipation is like adding blocks side by side in a sealed container full of water—the more blocks you add, the greater the water pressure between them. For human society, the more people enter, the more "crowded" it becomes. It was through this figurative thinking that I grasped the essence of rent dissipation—a cost that only exists within society.

[44] Traditional rent dissipation analysis indeed requires the large premise of countless entrants, but in the case of data goods, as long as there is sufficient demand and just two participants in competition, the rent can also be driven to zero. Stage 1, A, the sole provider in the market, sells an uncopyrighted digital course priced at 100

Apparently, rent dissipation is a macro (institutional) and meso (organizational) concept rather than a micro (individual) one. This is because the impact of rent dissipation affects collectives, transforming high-quality assets into mediocre ones. However, for any individual micro-level entity, their marginal benefit always equals their marginal cost, and their compensation will not fall below the societal average wage. Thus, the individual is always in the optimal state, even after rent dissipation has decreased to zero. Therefore, observing individuals has limited significance.

The connection between rent dissipation and transaction costs stems from Cheung's critique of the second formulation of Coase theorem. The premise of Coase theorem's second formulation is "property rights are clearly delineated" and "all costs of transactions are zero." Cheung argued that the assumptions of zero transaction costs and well-defined property rights are contradictory. This is because if transaction costs were truly zero, there would be no need for property rights or market constraints (Cheung, 2019a, pp.57-61). As we mentioned earlier, Coase agreed with this statement. Consequently, Cheung (2019a, p.62) acknowledged that "markets emerge because of transaction costs or, more broadly, institution costs". However, maintaining the existence of markets and property rights incurs costs. Under the assumption of self-maximization, individuals are willing to incur costs to uphold markets and property rights only to reduce the costs of the absence of market and property rights conditions. So, what are these social costs? Cheung claimed that these social costs are rent dissipation. He stated:

---

RMB. Stage 2, due to the absence of copyright restrictions, B purchases the course from A and resells it for 50 RMB. The clear price advantage causes A's original customers switch to B. Stage 3, seeing B's price reduction, A is forced to lower its price further. Stage 4, as demand in the market tends to infinity, both A and B's price reductions also tend to infinity, and ultimately, the price of the product approaches zero. This analysis demonstrates two points: First, the reason goods have a price is due to the existence of property rights; without property rights, there is no price. Second, this analysis explains why the price of pirated digital goods is extremely low. The reason why not all digital goods' prices approach zero is that rent is both a cost and a source of income. If the rent (price) were zero, the seller's income would also be zero, and the sales revenue would not even cover the selling costs, forcing the seller to exit the market, thereby ensuring a certain level of profit. However, for digital goods delivered by automated systems, since their operational costs are very low, this restriction does not apply.

> "Rent dissipation can only occur in the case of competition between individuals, so it can only occur in society. This dissipation is the cost that society needs to bear, so it is a cost, an expense. Rent dissipation is not directly related to production itself and is impossible in Crusoe economy. These characteristics are identical to transaction costs, but they can occur even without transactions, and it is more appropriate to call them institution costs in a broad sense. Broadly speaking, many other transaction costs can also occur without transactions, similar to rent dissipation, only in society." (Cheung, 2019a, p.62)

This assertion lays the groundwork for expanding the concept of transaction costs.

In summary, the essence of rent dissipation can be succinctly summarized: under unconstrained conditions, everyone pursues their own maximization, ultimately leading to a decrease in collective interests and the transformation of "better" into "mediocre." This tragic outcome is precisely the result of rational choices made by individuals.[45] Traditional economic theory aims to allow everyone to freely maximize their self-interest. However, rent dissipation theory suggests that without necessary constraints, even if everyone maximizes their own benefits, it does not mean that resources achieve optimal allocation.[46] Therefore, to reduce rent dissipation and achieve optimal resource allocation, appropriate competition rules, or institutions, must be established. In conclusion, transaction costs are not the costs of each individual maximizing their own interests but the costs incurred by everyone maximizing their own interests, making them the "side effects" of human society.[47]

---

[45] I am a supporter of human nature theory. I am aware that human civilization can shape people's character, but I am more aware that this progressive optimism cannot change the fact that "everyone has their essence". Each person carries their original sin, regardless of whether the starting point is Christian position. My inability to face my original sin does not mean it does not exist. The stronger the optimism under progressivism, the more hypocritical humanity becomes, deceiving itself more seriously, and the more morality decays. Someone once said that all revolutions since the French Revolution are just replicas of it. Here's the problem: social, historical, cultural, and economic conditions have undergone tremendous changes, yet the events that unfold remain the same. Why is this? This fully demonstrates that, at least in the past two hundred years, we have made no progress. Even if human nature may change one day, it has not changed since history was first recorded, and therefore I have no reason to believe it will change in the future. The only way to deny this point is to reject the premise of self-interest—that is, there is no complete self that allows humans to act in their own interest. Apparently, according to this standard, only certain types of psychopaths would be considered as non-self-interested. I am also aware that there is a group of people who claim that everyone has mentally disease, just to varying degrees. For these people, I only want to say that I respect common sense, and any person with common sense understands that someone without a medical qualification and clinical experience is not qualified to discuss whether another person is mentally ill, let alone to discuss whether all of humanity is mentally ill. In my view, these people are the epitome of learned foolishness. Selfishness is one of the most deeply rooted traits of human nature. Countless people have tried to challenge the dark side of human nature, but they have all failed without exception. However, they either refuse to admit that these disasters were caused by them or their spiritual mentors, or they blame it on humanity's self-interest. If history cannot be changed, then invent a new history—that kind of weakness is appalling.

As a result, Cheung reached the culmination of Coase's first perspective (task), explaining why proper constraints are necessary. While it is challenging to find instances of complete rent dissipation in real life, we can illustrate the impact of rent dissipation with a simplified example of "office politics": consider an organization where profit is the primary goal, and every employee aims for promotion. In this scenario, brown-nosing is the only path to promotion (the competition rule, or constraint). Each individual's way of maximizing themselves is to cater to the leadership as much as possible. While such behavior of pandering instead of productivity undermines the performance of the entire unit, it is the optimal choice for each employee to seek personal gain. Consequently, due to the inappropriate brown-nosing rule, the profitability of the organization must be weaker than that of units with performance-based competition rules. In this example, transaction costs (rent dissipation) are not the verbal flattery expended by each individual toward the leadership but the chaotic atmosphere resulting from everyone brown-nosing, leading to a decline in overall unit performance and diminished competitiveness for the enterprise.[48] For society, we could term this phenomenon "social muscular dystrophy," the terrifying consequence of which is a society where resources are wasted, individuals with ideas are constrained to conformity, and capable individuals are unable to leverage their skills, resulting in the squandering of all quality endowments and talents in meaningless internal struggles.

---

[46] In economic terms, it typically refers to maximizing total societal output.

[47] Civilized people are essentially as self-interested as savages. The difference is that each savage pursues their own maximum benefit, resulting in a decline in collective welfare; while each civilized person also pursues their own benefit, but ultimately leads to an increase in collective welfare. And this is the boundary between civilization and barbarism.

[48] Certainly, it is precisely because of employees' lack of productivity that the unit's performance declines. Therefore, the costs incurred by this lack of productivity are inseparable from the costs incurred by the decline in unit performance. Rent dissipation theory provides a new perspective on costs: for each individual, they are always in the optimal state, but for the collective, overall maximization is not achieved. In this example, the marginal cost and marginal benefit for employees are always equal, so individuals do not incur losses. The real losses are borne by the entire unit.

### 3.3 Institution Cost

Institution costs are a comprehensive concept— "because different categories of transaction costs are often inseparable, a broad interpretation is needed: transaction costs include all costs that are impossible in Crusoe economy's one-person world. In a one-person world, there is no society, nor are there economic institutions, so transaction costs become all costs that arise because of society—these can be termed institution costs" (Cheung,2019a, p.81). This is the first characteristic of institution costs; they are related to society and are broadly defined transaction costs. Specifically, transaction costs (institution costs) "includes not only the costs of forming and enforcing contracts (including seeking information for market transactions), but also the costs of delineating and policing exclusive rights (including those of institutional arrangements such as legislative enactments)" (Cheung,2006e, p.370). It is worth mentioning that Cheung believes that the term "transaction costs" does not accurately reflect its social nature, and "institution costs" should be used as a superior replacement for "transaction costs." However, since "transaction costs" have become a common term in academia, he retains this concept (Cheung, 2005, p.103).

There are various ways to constrain competition beyond market institutions. People queue up when buying goods, adhering to a first-come, first-served competition rule; in promotions, employees are ranked by seniority, following an age-based competition rule. Such constraints are ubiquitous, subtly influencing all aspects of social life. Without such rules constraining behavior, people would inevitably resort to conflict for personal gain, resulting in extensive rent dissipation. Faced with these realities, Cheung (2019a, p.152) realizes:

> "Similarly, all customs, ethics, religions, ceremonies, and laws imply constraints on competition, all of which can be regarded as contracts: explicit or implicit, voluntary or compulsory, all implying mutual agreement. ... Following this logic, the cost of constraining competition is the cost of contracts, and therefore also transaction costs, broadly termed as institution costs."

This is the second characteristic of institution costs; they are transaction costs from the perspective of contracts (constraining competition).

Cheung adopted the term "institution costs" to avoid the specific delineation required when using the concept of transaction costs and to meet the practical need for expanding its application scope. Consequently, the scope of transaction costs extends to the operation of society and becomes linked with contracts.

### 3.4 The law of transaction cost substitution[49]

The essence of transaction costs, rent dissipation, and institution costs is the same, but they also substitute for each other. This is known as the law of transaction cost substitution. Next, we will start our exploration of this law by considering another form of rent dissipation (also originating from Alchian's oral tradition) as our starting point.

---

[49] Let us now consider an ethical proposition: Would you prefer to live in a society where everyone is absolutely altruistic, or in a society where everyone is absolutely self-interested? Perhaps many would view the former as the "ideal world", while the latter represents a pit of material desire. If I had to choose, I would choose the latter. Let us approach this question using the analytical framework of "resource allocation-competition principles". What would the competition principles be in a society where everyone is absolutely self-interested? If everyone is self-interested, it means that people always consider what others can offer them before making a choice. In other words, people would allocate resources to whoever is willing to pay the highest price, which aligns with the "highest bidder wins" market mentality. Of course, issues such as violence and factionalism may arise, but since we are merely conducting a thought experiment, let us put those aside for now. Ultimately, this would result in a society that is highly competitive, favors the strong, and ensures that resources are largely utilized. Now, what would the competition principles be in a society where everyone is absolutely altruistic? If everyone is altruistic, it means that people always consider who is most in need of the resource at the moment. In other words, those with the most urgent need for a particular resource would receive it from others. However, the most urgent human desire is simply survival. For example, when drowning, a person may unconsciously drag the one trying to save them under the water. Survival is a basic instinct, and the right to life is the most fundamental human right, far surpassing the concepts of altruism or self-interest. In such a society, all resources would be used solely to maintain basic survival, and everyone would be in a state of barely existing. The result would be an extremely weakened society. Now, which society would you prefer to live in?

Resources are scarce. As economists, we aspire for the most capable individuals to obtain and utilize resources, inevitably introducing competition. However, the theory of rent dissipation tells us that competition must be constrained, or else it leads to the tragic outcome of everyone being exhausted, and society as a whole suffering losses. Therefore, where there is competition, there must be rules, and those who adhere to the rules more closely prevail. From this perspective, we can extract two key elements—institutional purpose and Competition criteria. Ideally, institutional purpose and competitive criteria align, with competition acting as a filter to derive resource allocation patterns most in line with institutional goals. However, these two are often not perfectly aligned (Cheung, 2019b, pp.108-113).

Let's consider the example of the examination system. Suppose the purpose of the examination system is "to allow outstanding individuals to access high-quality educational resources." However, a subsequent question arises: how do we measure excellence? Clearly, we cannot directly judge whether someone is "excellence." Fortunately, there is a certain correlation between exam scores and excellence, so we set the competitive criterion as exam scores—higher scores imply greater "excellence." However, scores can only partially reflect a student's abilities, leading to issues of conversion rates. Different systems have different conversion rates, which determine system performance. Under an examination-oriented system, people's rational choice is to strive to improve their exam scores. As a result, outstanding individuals expend effort to prove their excellence, while less outstanding individuals, adept at test-taking, are promoted. Although the examination system aims to select the most "excellence" talents, it ultimately selects the students most skilled at test-taking. While test-taking ability is related to excellence, it does not equate to "excellence", leading to the waste and misallocation of resources. Nonetheless, it is essential to note that while the examination system may not be perfect, its dissipative level is certainly lower than that of a recommendation letter system. This is why we ultimately choose the former over the latter.

In summary, deviations between institutional objectives and competition criteria also lead to rent dissipation. Cheung (2019b, p.112) provided several examples: "Using age as a criterion for distribution encourages individuals to spend money and effort on falsifying their age or increasing their desire to age quickly. In a society where might makes right, investing in weapons is encouraged." These resources, which could have been allocated to increasing output or other activities directly related to institutional objectives, are ultimately consumed during the competition process. Cheung clearly pointed out that to eradicate rent dissipation, only price should serve as the competition criterion (Cheung, 2019a, p.71). Specifically, this assertion should be "only the price mechanism prevents deviations between institutional objectives and competition criteria", where the institutional objective is to increase output, and the competition criterion is who produces the most, thus ensuring no deviation. Figure 2 visually represents all the analyses above.

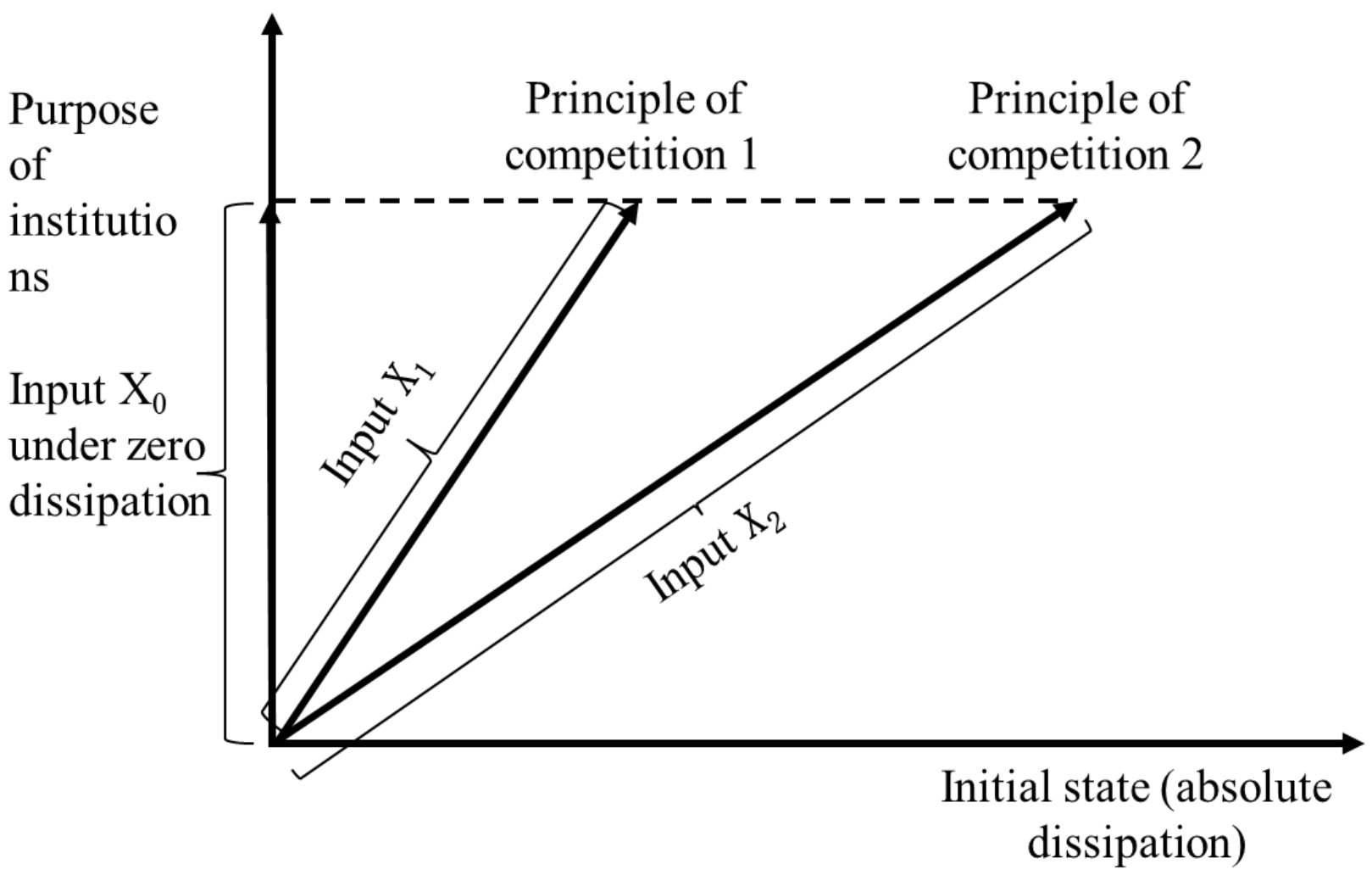

**Fig. 2** Analysis of rent dissipation from the perspective of the Law of Substitution of Transaction Costs

Given this, why do other forms of competition criteria still exist in the world? Why do people prefer to incur the substantial cost of rent dissipation rather than choosing a price system? The answer is that adopting a price system also incurs costs (institution costs), which are greater than the losses caused by rent dissipation. Similarly, people choose price systems for certain issues because the costs of using a price system (institution costs) are lower than the losses caused by rent dissipation (Cheung, 2019a, pp.72-73). This is an example of the mutual substitution between institution costs and rent dissipation. In conclusion, Cheung uses the transaction cost substitution theorem to describe the connection between transaction costs, institution costs, and rent dissipation, thereby explaining the choice of contracts and institutions.[50]

---

[50] By this point, readers may feel that competition principles are good thing, for it can prevent rent dissipation.

## 4. The New "Trinity"[51]

### 4.1 Unresolved Issue: Contradictions in Coase's Two Perspectives

The two perspectives on transaction costs have reached an impasse. The essence of Coase's first perspective lies in rent dissipation, while the second perspective centers on constraint conditions. However, conceptually, they are in conflict. As previously discussed, the first perspective refers to the costs incurred when each individual maximizes their own utility, whereas the second perspective views transaction costs as the costs individuals incur to maximize their utility (in part). On one hand, transaction costs are the costs of maximizing one's utility, yet on the other hand, they are not. This contradiction is apparent.

The contradiction between the two perspectives is specifically reflected in the concepts of rent dissipation and institution costs. As mentioned earlier, Cheung's description of the relationship

However, I will now reveal the harsh side of competition principles (institutions) —the so-called competition principles are merely tools designed to prevent people from fighting (avoiding rent dissipation). Bloodshed is painful, but competing under foolish principles is equally unbearable. In China, in 2014, university students could easily find good jobs. But by 2024, without a graduate degree, you might not even get a job. If you're a capable undergraduate, your options are: (1) Waste the best three years of your life on meaningless graduate courses for a chance at a potential good job; (2) Accept unreasonable degree-based discrimination and lower salaries. Graduate students are not much smarter than undergraduates; we all know what graduate education in our country is about. But employment discrimination is real. It is illogical and unjustifiable, but it exists simply to prevent people from fighting each other for jobs. Here is the brutal side of competition principles — theoretically, systems are meant to reduce rent dissipation and improve society. However, their purpose is not to improve society, but merely to prevent people from fighting. Institutional costs and changes follow not a mechanism of "choosing the best", but "who is the least bad". Let me make this clearer: Mr. Wang has two fierce dogs. They fight every day over the same piece of meat, making Wang awake at night, but he can't afford a second piece. So, Wang sets a rule: only the dog that can blow the biggest snot bubble gets to eat the meat. After that, the two dogs never fought again, and spent their time figuring out how to blow bigger snot bubbles. And this competition could be who can make the stinkiest poop, who can urinate the furthest, or who can pass the most pungent gas. As long as no one dies, people will tolerate even the most absurd rules. And as long as it's better than the worst, it won't change. Do you understand now?

[51] For the first two chapters, some people have made comments. But in the third quarter, many people could not understand it. The fault lies with me, because I didn't write clearly. Before the "New Trinity", I owe everything to Cheung, whether it was my own idea or Cheung's. After the "New Trinity", whether Cheung has said it or not, it all belongs to me.

between transaction costs, institution costs, and rent dissipation is approached from "three different perspectives on social costs." In other words, all three ultimately pertain to social costs, differing only in quantity rather than quality. However, the problem lies therein. Rent dissipation represents the primal state of human society, with the logic being that under completely unconstrained conditions,[52]rent dissipation constitutes the social cost of resource allocation. Institution costs, on the other hand, result from human intervention, based on the logic that there exists an optimal allocation state for resources, and the key issue is how to achieve this with minimal costs. It is evident that the difference between the two lies in whether people are actively involved. Additionally, rent dissipation cannot be used as a micro-level concept, whereas institution costs, as an extension of transaction costs, can be applied as a micro-level concept, with the two having fundamentally different subjects of application.

In conclusion, the nature of these two concepts is certainly different. By distinguishing between rent dissipation and institution costs, we anchor the two perspectives of transaction costs to two distinct concepts. We associate rent dissipation with the first perspective and institution costs with the second perspective. The question now is: which one of these is the transaction cost we need?

### 4.2 The New Law of Transaction Cost Substitution: A Qualitative Analysis

The essence of transaction costs lies in the shadows of obscurity, awaiting discovery through keen insight—how do institution costs substitute for the dissipation of rent? A careful examination reveals that Cheung equates institution costs with rent dissipation through the bridge of "contractual costs". As to why contracts are introduced, Cheung (2019a, pp.19-20) stated:

> "Without introducing contract analysis, the limitations of transaction costs are handled ambiguously, preventing a thorough analysis of resource utilization, income distribution, unemployment, inflation, and other phenomena... The laboratory of economics is the real world...

[52] It needs to be clarified that there is no absolute sense of "unconstraint"; "competition" itself constitutes a form of constraint. However, since competition is assumed to be absolute, it is considered an underlying assumption and axiom in economics, typically requiring no further discussion. Absolute rent dissipation is exceedingly rare; therefore, in practical application, a certain condition can be anchored as a benchmark state, referred to as a state of "rent dissipation," inspired by Cheung's perspective (see Cheung, 2005, pp.371-372).

> We need to examine on-site what the actual verification conditions are...... Such examination is never easy...... But assuming the limitations of the real world in this way is close to boring nonsense. A compromise is to first explain contract selection with transaction costs and then explain the phenomena or behaviors of resource utilization and income distribution from the constraints of contracts."

It is apparent that Cheung also opposes overly detailed categorization of transaction costs; instead, he attempts to construct a system of "transaction costs → contracts → phenomena". The above argument seems to stem from the logic that there naturally exist many choices "parallel" to the initial state (rent dissipation), corresponding to different amounts of costs (institution costs), and people only need to instinctively choose the option with the lowest cost to minimize costs. When explaining rent dissipation, Cheung (2019a, p.152) wrote: "Not all constraints in human competition need to be viewed from the perspective of contracts; however, it is an optional viewpoint. Sometimes, adopting a contractual perspective adds complexity, which is undesirable, but at other times, it clarifies issues." Integrating the above points, we find that Cheung only points out the necessity of introducing contracts but still leaves some unanswered questions—why can people view things from the perspective of contracts? Why is it sometimes more convenient to view things from the perspective of contracts? What are the differences between contracts and the real world? We hope to discover the process of institution costs substituting for rent dissipation in these questions.

The real world is inherently complex, so emphasizing only "costs of negotiation, monitoring, etc." is inaccurate—more precisely, it is "due to the complexity of real-world conditions, individuals cannot directly constrain behavior, thus incurring high costs." Because of this, as in the example by Cooter and Ulen, people are compelled to argue the unarguable:

> "Suppose that A contracts with B to play Hamlet in a production of Shakespeare's play at B's theater. Subsequently 4 breaches the contract by refusing to perform his role. Typically, a court will not grant B specific performance, that is, an order compelling A to perform Hamlet. This is because the costs to the court of judging whether A had discharged his contractual obligation to B

> are extraordinarily high. How should the court assure the quality of A's performance as the Prince of Denmark? Perhaps because A is in such a pique about his dispute with B that without stringent and expensive supervision by the court he will seek to embarrass B by the shoddiness of his performance as Hamlet. When these supervisory costs are recognized, it may be more efficient for the court to award B money damages." (Cooter & Ulen,1988, p.324)

The issue B faces is ensuring A's full performance capability. However, I's evident that straightforwardly stating B's goal in the contract without adjustments would not aid its enforcement or monitoring. The dilemma lies in the complexity of the situation, where B struggles to find effective monitoring and measurement methods, leading to prohibitively high costs, making it impossible to either sign or, if signed, inevitably damaging to one party. As observers, we might offer practical suggestions to B, but we have no reason to assume that B has the ingenuity to think of these methods—which is precisely the purpose of consulting firms.

Let's consider another example from Umbeck. Unlike the previous case, a solution was found here—people sought to address smoke pollution caused by fires but ultimately opted to restrict chimney construction—"For example, if my neighbor builds a fire in his fireplace during the day, the smoke that dirties my air can be seen and appropriate measures taken for the enforcement of my property. If the fire is burned at night, the smoke is more difficult to detect, and theft more likely to occur. Any action involves the use of more than one resource, some of which are more easily detected than others. In the fire example, the use of air to disperse the smoke may be very costly to detect. But the burning of an indoor fire usually requires the use of a fireplace and chimney which are easily observed. It is here that one of the economic justifications of restrictions on exclusive property is found. By restricting homeowners from building fireplaces, the costs of enforcing property rights to air can be reduced" (1981, pp.113-114).

These two examples highlight an important yet often overlooked issue—contracts embody people's problem-solving thinking, yet the complexity of reality often leads to a certain deviation between contractual requirements and the contract's purpose. For instance, the examination system aims to select the most "excellent" talents, but because "excellence" cannot be directly measured, it ultimately selects the students who are best at taking tests. We must explain the reason for this deviation. The answer is clear: the deviation arises not because people intentionally choose suboptimal rather than optimal solutions, but because people designing the contract are subject to certain constraints; in other words, they cannot find better solutions. Indeed, people's choices of contracts are constrained by transaction costs (institution costs) that enable the contract to function, meaning they tend to choose contracts with the lowest implementation costs. However, it is equally important to recognize that the process of finding these solutions, corresponding to different costs, is also constrained. This implies that although contracts attempt to reflect and address real-world problems, there is an unbridgeable gap between them—not a matter of cost magnitude but of ability. At this point, statements such as "the cost of drafting contracts is extremely high" related to "quantity" merely evade the distinction in "quality". We need to return to the fundamental condition of "complex reality" and answer the following questions: Where do these contracts corresponding to different costs come from? More precisely, what is the process from complex reality to a written contract?

Each individual seeks to maximize within certain constraints; conversely, solving real-world problems is subject to specific constraints. These constraints are myriad, some easy to handle while others are not. Therefore, the key lies in identifying which constraints are relatively important and which are relatively unimportant, a process referred to as simplification of the real world. In fact, it's not only theorists who attempt to simplify the world; everyone does so, or else progress would be impeded. However, some individuals excel at simplification while others have mediocre abilities in this regard. Contracts embody the manifestation of simplification of unorganized competition, as individuals exert powerful constraints by controlling key points.[53] Cheung's shortcoming lies in his belief that simplification exists only in the theoretical analyses of economists, overlooking the simplification undertaken by each individual. Thus, we arrive at the answers to the initial three questions—contracts

[53] It needs to be pointed out that contracts possess a unique nature. On one hand, they embody simplification by individuals, while on the other hand, they themselves are a form of simplification by individuals.

embody people's simplification of the complex world.

"Perceive the subtle to grasp the significant."[54] Institution costs are not separate; they are precisely the simplification of rent dissipation, encompassing not only economists' simplifications but also those made by individuals in real-life situations. In this regard, predecessors have made significant strides. For instance, Williamson emphasizes the impact of asset specificity, while Barzel categorizes various forms of costs under information costs (though this approach is debatable); these are economists' simplifications. Conversely, the emergence and refinement of private property rights represent simplifications made by people in social evolution. Consequently, the genesis of institution costs has a real basis and should not blindly be labeled as "hindrances". Fundamentally, the ultimate purpose of institution costs is to constrain competition; however, at times, they may work contrary to intentions or target the wrong entities. Hence, institution costs are neutral—when simplified correctly, they can establish reasonable competition standards and enhance social welfare; when simplified incorrectly, they may establish inappropriate competition standards, further reducing collective welfare. Thus, two concepts and two perspectives are unified—rent dissipation corresponds to the first perspective, representing the costs incurred by individuals maximizing their own interests, while institution costs correspond to the second perspective, representing the costs of reducing social costs by simplifying rent dissipation into several dimensions. Naturally, these are costs incurred in maximizing one's own interests, or what economists can observe as "real factors".

[54] The previous several paragraphs are all redundant and were not originally included. I added them because I realized no one understood my point, so I filled in the thought process afterward in hopes of clarifying. The reason for the lack of clarity is that I had to use academic language to express my ideas, rather than the plain language I'm using now. By the way, simplifying academic language into plain speech is the most basic skill for a scholar. If I had followed my own thinking pattern, I would have skipped straight to the "simplification" part. To summarize this section in one sentence: Some people design contracts (institutions) that are simple and easy to implement, while others design contracts (institutions) that are chaotic and inefficient; The difference isn't about some fictitious "execution costs" — it's simply because the people who designed the contracts (institutions) are too inept to come up with good design methods; The difference between good and bad design methods lies in whether the core key points are identified, and the process of identifying those points is the process of simplification; In any case, we all need to be smart, to focus on the key points, and to learn to simplify. If we translate the above analysis into the "institutional cost—rent dissipation" framework: what has been simplified is the institution, and the operation of the institution has a cost, known as institutional cost. The object of simplification is the chaotic reality, full of trivial disputes for small gains, which is rent dissipation. Though both rent dissipation and institutional cost are societal costs, institutional cost involves active human participation and is a simplification of rent dissipation, so I have successfully distinguished the two and reconstructed the transaction cost substitution principle.

Without the concept of rent dissipation, everyone pursues their own maximization, and each person's actions are deemed correct and rational, with no notions of "obstacle" or "hindrances". In this context, institution costs are merely a generic term for a portion of the costs incurred in maximizing outcomes, possessing only nominal value without economic significance. At this point, institution costs are simply a generic term for part of the costs incurred in achieving maximization, possessing only nominal value and lacking economic significance. Therefore, introducing the concept of rent dissipation is essentially introducing a scale to measure human interactions—if institution costs are either positive or negative, then rent dissipation represents that miraculous and profound zero. The error of traditional transaction cost theory lies in its failure to discover rent dissipation as a "side effect" of human society. Additionally, it mistakenly treats the remedy prescribed by humans for themselves—institution costs—as the ailment itself (the impediment). Rent dissipation serves as an anchor for assessing the value of institution costs, which, as costs constraining competition, reflect the efforts to resolve rent dissipation—people first simplify, then substitute one type of cost (the cost incurred for personal gain) for another (the cost each incurs for their own gain).

In summary, a major task of transaction cost economics should be to investigate how the pursuit of individual maximization can diminish collective welfare, and to discern which institution costs are justified and which are not. In terms of institution design, the true subject worthy of exploration and study is indeed institution costs, but it is essential to introduce the concept of rent dissipation to highlight their value. At this point, transaction costs merely serve as a collective term for both, possessing only a nominal descriptive value. It should also be noted that in Marxist political economy, the concept that stands in opposition to transaction costs is not circulation expenses, but rather the organization of production by capitalists and planned economies. It should be noted that the analysis above does not change the fundamental nature of both rent dissipation and institution costs as

"constraint condition."

### 4.3 The New Transaction Cost Substitution Law: Quantitative Analysis

The primary objective of this section is to elucidate the substitution relationship between institution costs and rent dissipation more accurately through quantitative analysis. Imagine an independent community in its original state; we will analyze it in four steps to understand how a portion of the actual institution costs incurred might, in fact, represent savings in rent dissipation. In other words, what is traditionally considered a societal cost becomes a saving on societal costs. Furthermore, we will recognize that quantitative substitution, similar to qualitative, is not a straightforward replacement.

Step One corresponds to the first row in Figure 3. In this community, there are no types of contracts or institutions, and competition is disorderly—this represents the most primitive state of resource allocation in society. We assume that under these conditions, rent dissipation is quantified as 1.

Step Two corresponds to the second row in Figure 3. At this point, people realize that excessive competition leads to a decline in collective welfare. Consequently, they elect a distributor tasked with maintaining the state of resource allocation with minimized rent dissipation, compensating him with a value of $a$. The removal of one person from competition results in a decrease in rent dissipation. For the sake of discussion, let us assume that the community has a large population. Therefore, the reduction in the amount of rent dissipation is minimal, approximating zero. Hence, the institution cost is $a$, and the society saves an amount valued at $1-a$.

Step Three corresponds to the third row in Figure 3. The distributor's powers of enforcement and measurement are limited, and individuals, acting in their own self-interest, engage in maneuvers with the enforcer, thereby creating various new problems. These troubles arise from unclear definitions of income rights and thus constitute a new form of rent dissipation, valued at $b$. Consequently, the institution cost remains $a$, and the rent dissipation is $b$. The actual savings for society costs are now valued at $1-a-b$, rather than the original $1-a$.

Step Four corresponds to the fourth row in Figure 3. The first three steps overlook an issue: although the distributor is elected by the community, he also has the right to refuse the role, otherwise the maximization of everyone else's interests comes at the expense of the distributor's individual welfare. Assume that the distributor was originally a fisherman, earning an income equal to the societal average, denoted by $c$. He agrees to take on the role because the community offers him higher wages and he is capable of performing the duties (for instance, he may be highly respected and able to command authority). The community is willing to pay him this income because his work improves their collective welfare. Since this is a cost of maximizing collective benefits, it should not be considered a societal "burden." Since a new role of distributor has been added to society, there is one less fisherman earning $c$, but there is now a distributor earning $a$. The distributor's income has increased by $a$-$c$, which should also be considered a saving in rent dissipation. Therefore, the institution cost is $c$, rent dissipation remains b, and the actual savings for society costs are valued at $1$-$b$-$c$, not $1$-$a$-$b$.

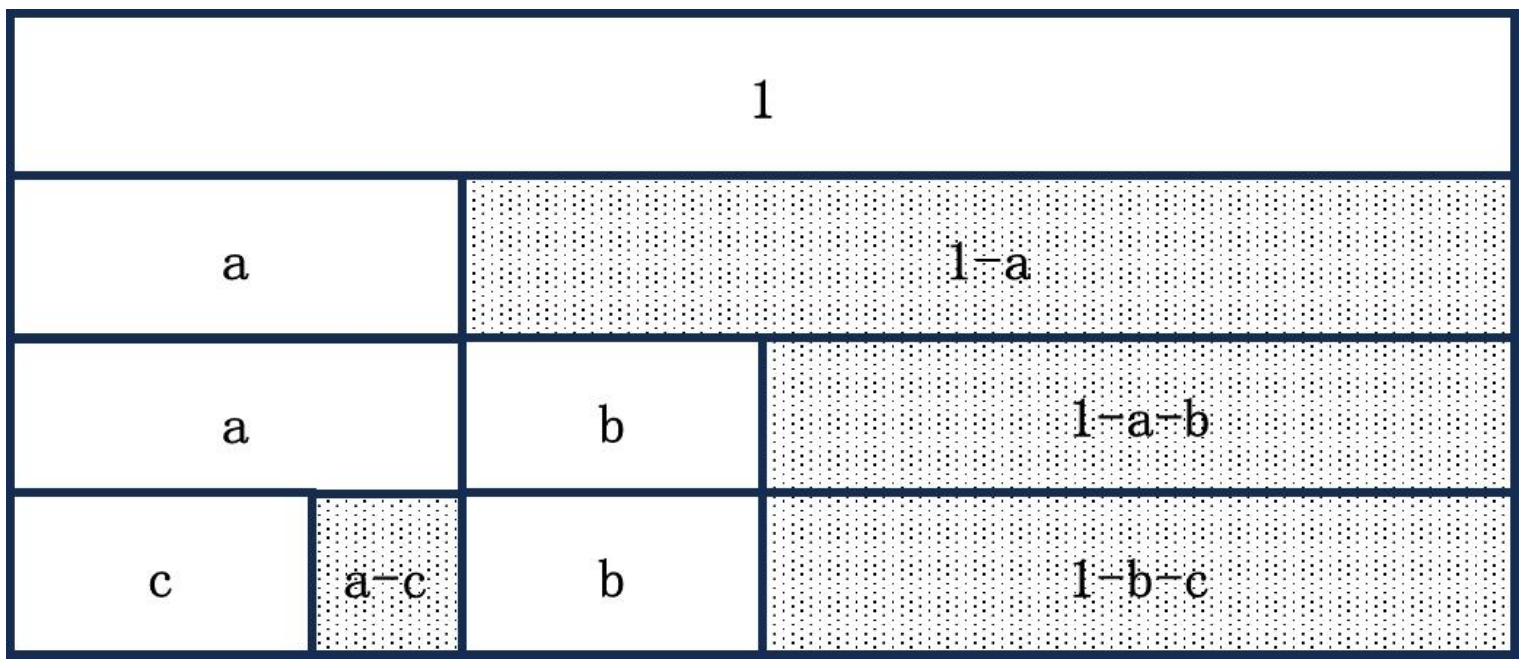

**Fig. 3** Quantitative analysis

From this analysis, we can draw two conclusions: (1) Institutions (constraints) inevitably lead to new dissipations of rent, so until a perfect system emerges, the costs of instituting such constraints can only partially substitute for this dissipation of rent. (2) The expenses for maintaining institution operations are not necessarily a social burden; rather, they might represent savings from rent dissipation. In principle, the specific amount equals the difference between the actual revenue and the societal average revenue (or the highest income that could be earned independently of institution costs). This analysis is entirely based on the principles of opportunity cost and comparative advantage.

The above analysis aims to demonstrate that the costs associated with institutions that resolve rent dissipation are not a societal burden but rather an integral part of societal harmony. This view is entirely contrary to traditional theory; however, it becomes more acceptable when considered from the perspective of real life. Consider the example of an intermediary who provides information to both parties in a transaction, effectively preventing fraud and generating benefits for both, thereby making their service indispensable. Traditional views hold that the intermediary's commission is an institution cost (transaction cost), and that higher institution costs correlate with lower economic performance. In other words, the income of intermediaries, who provide services to the market, is seen as an impediment to economic development—the lower their income, the higher the economic efficiency. Clearly, this conclusion is unacceptable. According to the perspective of this paper, in principle, the part of the intermediary's current income that exceeds the highest income they could earn independently of institution costs and rent dissipation represents the absolute savings from rent dissipation. Regarding the measurement of the total transaction costs, the analysis shows that simply aggregating the costs incurred in transactions, legal processes, and institutions is meaningless. Instead, the crucial factor is the difference from the original state.

It is important to note that from a quantitative perspective, institution costs include not only the necessary expenditures to maintain the operation of systems but also the misallocation of resources that institutions cause. Consider the example of traffic flow at an intersection. On both roads, each driver tries to maximize their speed. Without traffic lights or a designated person to coordinate, the outcomes would either be extremely slow speeds or frequent accidents. This would transform a highway intended for fast vehicular passage into a road plagued by hazards, resulting in a decline in its value (rent dissipation). Suppose people choose to install traffic lights to address this issue; the cost of purchasing and maintaining these traffic lights represents the traditional concept of institution costs. While the traffic light system is designed to alleviate congestion, it also leads to resource waste when drivers must stop at red lights even when there are no other vehicles. Therefore, this loss should also be accounted for in the calculation of institution costs.

## 5.Institution Costs and Society

## 5.1 Social Costs Are Not Merely the Sum of Individual Costs

Rent dissipation reflects atomistic interactions among individuals, each striving to maximize their own interests, thereby generating a collective force. This dissipation of rent is the social cost associated with creating this collective force. By simplifying the process of forming this collective force, we can achieve the same or better results at a lower cost, indicating that the total social cost is not merely a straightforward aggregation of individual costs.

Laws are effective because the power of an individual is insignificant compared to the force of law. Similarly, the threat or deterrent effect of violent groups exists because the power of an individual is negligible against a more formidable collective. However, there is no reason to assume that social costs are always less than the sum of individual costs. In fact, the idea of simplifying institution costs is a double-edged sword. For those committed to promoting social progress, this approach reduces social costs, but for those with ulterior motives, it involves lowering localized costs for profit, which in turn increases the overall social cost.

## 5.2 Two Forms of Institution Costs

The issue of institution costs fundamentally concerns the choice of governance. In modern societies, this boils down to rule by law versus rule by man. The former operates on principles, while the latter relies on power. However, this does not imply that there is no power in societies governed by law or no law in societies governed by a single individual. The substitution and complementarity between the two are also constrained by the law of transaction cost substitution.

Institution costs manifest in the forms of principles and power. Principles are explicit and consistent, whereas power is adaptive and elusive. A society operating on principles implies a minimal existence of exceptions; on the other hand, a society operating on power is characterized by a pervasive practice of making exceptions for special cases. Under principles, people act freely within established rules; under power, those in authority use their leverage to influence society, making others merely extensions of their less responsive limbs.

Incorporating principles into economic analysis is relatively simple, but integrating power into economic analysis is significantly more complex. Economists are familiar with the economics of resource allocation based on price mechanism; the next step should be to study the economics of resource allocation based on power.

### 5.3 Self-Reinforcement of Institution Costs

Institution costs inevitably lead to rent dissipation. To address this, people may choose to invest in new institution costs, which in turn inevitably leads to further rent dissipation. The result of multiple iterations could be that the total amount of institution costs exceeds the amount of rent dissipation in the original state. Therefore, if there are fundamental deficiencies in institutional design, the societal burden will not decrease due to the establishment of institutions; rather, it will continue to accumulate as the system evolves. Without a healthy exit mechanism, organizations and institutions can create “tumors” for themselves in their daily operations, leading to a continuous accumulation of total institution costs in society.

### 5.4 Institution Costs and Economic Performance

When evaluating the performance of institutions, it is essential to analyze them in conjunction with their intended purposes. Economists tend to focus on how institutions affect the maximization of output. However, often the primary purpose of institutions is not necessarily to enhance social welfare and economic output but to maintain social stability, cater to special groups, or efficiently extract rent, among other goals. For institutions with a notorious reputation, economists often label them as “inefficient” due to their tendency to decrease output, attributing this to the extremely high institution costs. But if the goal of an institution is not to maximize output but something else, then such criticism by economists is undoubtedly a misjudgment.

Therefore, the “error” in institutional purposes (values) should not be solely attributed to high institution costs. A typical example is the military merit system implemented in the Qin dynasty. This system bound the people of Qin to the wars of unifying the six states, leading them to prioritize killing enemies over agricultural production. This significantly intensified the intensity of warfare and undoubtedly resulted in a decrease in total societal output. This significantly intensified the intensity of warfare and undoubtedly resulted in a decrease in total societal output. If we were to examine it from the perspective of output maximization today, no economist would refrain from criticizing the reforms implemented by Shang Yang. However, it cannot be denied that this system was highly effective in strengthening the Qin military and was an essential institutional guarantee for unification. Therefore, it is not appropriate to simply label it as "inefficient." The correct approach is to consider whether it is possible to overcome certain limitations within the institution, find better ways to simplify it, and seek new avenues for improvement.

Furthermore, when evaluating institutional performance, one must not overlook the benefits brought about by the institution. As mentioned earlier, the logic of institution costs lies in achieving the institution's objectives (often conceptualized by economists as output maximization) at minimal cost. However, there is no reason to believe that higher institution costs lead to lower institutional performance. Ignoring the measurement of institutional benefits while only calculating institution costs is akin to turning a blind eye to the bigger picture.